\documentclass[traditabstract]{aa} 

\usepackage{graphicx, times, txfonts, float, rotating, color, url, lscape, bm} 
\usepackage[normalem]{ulem}

\newcommand{\less}{\raisebox{-1.1mm}{$\stackrel{<}{\sim}$}} 
\newcommand{\more}{\raisebox{-1.1mm}{$\stackrel{>}{\sim}$}}

\newcommand{\mum}{$\mu$m} 
\newcommand{\muas}{$\mu$as} 

\newcommand{\G}{{\it Gaia}}
\newcommand{\Hp}{{\it Hipparcos}}

\begin{document}

\title{The parallax zero point offset from \it Gaia \rm EDR3 data
\thanks{
Table~\ref{Tab-Targets} is available in electronic form at the CDS via 
anonymous ftp to cdsarc.u-strasbg.fr (130.79.128.5) or via 
http://cdsweb.u-strasbg.fr/cgi-bin/qcat?J/A+A/.
%Additional files are available via *** TBD ***
%Table~\ref{App:Fit} and Figure~\ref{Fig:App} are available in the on-line edition of A\&A. 
}
%\fnmsep
%\thanks{
%Based on observations 
%} 
}  
 
\author{ 
M.~A.~T.~Groenewegen
}

\institute{ 
Koninklijke Sterrenwacht van Belgi\"e, Ringlaan 3, B--1180 Brussels, Belgium \\ \email{martin.groenewegen@oma.be}
} 
 
\date{received: ** 2021, accepted: * 2021} 
 
\offprints{Martin Groenewegen} 
 
%\authorrunning{Groenewegen et al.} 
%\titlerunning{CO rotational line-emission in AGB stars in the LMC} 
 
\abstract
{
  The second data release of \it Gaia \rm revealed a parallax zero point offset of $-0.029$~mas based on quasars.
  The value depended on the position on the sky, and also likely on magnitude and colour.
  The offset and its dependence on other parameters inhibited an improvement in the local
  distance scale using e.g. the Cepheid and RR Lyrae period-luminosity relations.
  Analysis of the recent \it Gaia \rm Early Data Release 3 (EDR3) reveals a  mean parallax zero point offset of $-0.021$~mas
  based on quasars. 
  The \it Gaia \rm team addresses the parallax zero point offset in detail and
  proposes a recipe to correct for it, based on ecliptic latitude, $G$-band magnitude, and colour information.
  This paper is a completely independent investigation into this issue focussing on the
  spatial dependence of the correction based on quasars and the magnitude dependence based on wide binaries. The spatial and magnitude
  corrections are connected to each other in the overlap region between $17 < G < 19$. The spatial correction is presented at several
  spatial resolutions based on the HEALPix formalism. The colour dependence of the parallax offset is unclear and in any case secondary
  to the spatial and magnitude dependence.
  The spatial and magnitude corrections are applied to two samples of
  brighter sources, namely a sample of $\sim$100 stars with independent trigonometric parallax measurements from \it HST \rm data, and a
  sample of 75 classical cepheids using photometric parallaxes.
  The mean offset between the observed GEDR3 parallax  and the independent trigonometric parallax (excluding outliers) is
  about $-39$~\muas, and after applying the correction it is
  consistent with being zero. For the classical cepheid sample it is suggested that the photometric parallaxes may be underestimated by about 5\%.
}  


\maketitle

\section{Introduction}
\label{S-Int}

Data from the {\it Gaia} mission \citep{GC2016a} has impacted most areas in astronomy.
One the fields were the {\it Gaia} 2nd data release (GDR2, \citealt{GDR2Sum}) was eagerly awaited was in
reliably establishing the local distance scale through calibration of the period-luminosity ($PL$) relation
of classical cepheids (CCs) and RR Lyrae (RRL) variables.

\citet{RiessGDR2} analysed a sample of 50 CCs. They derived a parallax zero point offset (hereafter PZPO) of $-0.046 \pm 0.013$~mas,
compared to the $-0.029$~mas derived for quasars by \citet{Lindegren18} and concluded that the need to independently
determine the  PZPO largely countered the higher accuracy of the parallaxes in determining
an improved zero point of the $PL$-relation.
Independently, \citet{Gr_GDR2} (hereafter G18) derived a PZPO of $-0.049 \pm 0.018$~mas based
on a comparison of nine CCs with the best non-\G\ parallaxes (mostly from {\it HST} data).
\citet{Ripepi_GDR2} re-classified all 2116 stars reported by \citet{Clementini19} to be Cepheids in the Milky Way (MW).
Period-Wesenheit relations in the {\it Gaia} bands were presented.
Assuming a canonical distance modulus to the LMC of 18.50, a {\it Gaia} PZPO
of $\sim -0.07$ to $-0.1$~mas was found.
PZPOs based on GDR2 data were also reported for RRL stars ($\sim-0.056$~mas, \citealt{Muraveva18};
$-0.042 \pm 0.013$~mas, \citealt{Layden19}), and many other classes of objects
(\citealt{Stassun18}, \citealt{Graczyk19}, \citealt{Xu19}, \citealt{Schonrich19}).
These values were mostly all-sky averages, but when sufficient data was available it was clear that the PZPO
depended on position in the sky, magnitude, and colour \citep{Zinn18, Khan19, LeungBovy19, ChanBovy19}.

The {\it Gaia} Early Data Release 3 (GEDR3) presents the most recent information on parallax, proper motions, position
and colour information for about 1.8 billion objects \citep{GEDR3_Brown}. 
\citet{GEDR3_LindegrenAS} presents the general properties of the  astrometric solution,
while \citet{GEDR3_LindegrenZP} (hereafter L20) specifically addresses the PZPO and make a python script available
to the community in order to calculate the PZPO.
This module gives PZPO (without an error bar) as a function of input parameters ecliptic latitude ($\beta$),
$G$-band magnitude, the {\tt astrometric\_params\_solved} parameter \citep{GEDR3_LindegrenAS}, and either
the effective wavenumber of the source used in the astrometric solution ($\nu_{\rm eff}$, {\tt nu\_eff\_used\_in\_astrometry} for the
5-parameter solution {\tt astrometric\_params\_solved} = 31) or the
astrometrically estimated pseudo colour of the source  ({\tt pseudocolour)} for the 6-parameter solution ({\tt astrometric\_params\_solved}= 95).
The module is defined in the range 6 $< G <$ 21~mag, 1.72 $>$ $\nu_{\rm eff}$ $>$ 1.24~\mum$^{-1}$, corresponding to about
0.15 $< (G_\text{BP} - G_\text{RP}) <$ 3.0~mag where $G, G_\text{BP}$, and $G_\text{RP}$ are the magnitudes in the {\it Gaia} G-, Bp-, and Rp-band, respectively.

Several papers have already applied the L20 correction to the raw GEDR3 parallaxes.
\cite{RiessGEDR3} applied it to a sample of 75 CCs in the Milky Way (6.1~\less~$G$~\less~11.2~mag,
0.9~\less~$(G_\text{BP} - G_\text{RP})$~\less~2.5~mag).
For this sample, there is a strong dependence of the correction on $\beta$ (ranging from $-4$ to $-38$~\muas)
with only a small dependence (of order $~1.8$~\muas) on magnitude and colour, with a median correction of $-24$~\muas.
Allowing for a remaining PZPO after application of the L20 correction and fitting the data to the independently
calibrated PL relation of CCs in the Large Magellanic Cloud (LMC) reveals an offset of $14 \pm 6$~\muas, in the
sense that the L20 corrections are too much negative, that is the values are over corrected.

A similar conclusion is reached by \citet{Zinn21} who analysed a sample of 2000 first-ascent red-giant branch stars with asteroseismic
parallaxes in the {\it Kepler} field (9.0~\less~$G$~\less~13.0~mag, 1.0~\less~$(G_\text{BP} - G_\text{RP})$~\less~2.3~mag) and concludes that 
the L20 corrections are too much negative by $15 \pm 3$~\muas\ for $G$ \less 10.8~mag.

\citet{Bhardwaj20} apply a theoretical period-luminosity-metallicity relation in the $K-$band to a sample of about 350 Milky Way
RR Lyrae stars (8.9~\less~$G$~\less~17.8~mag, 0.4~\less~$(G_\text{BP} - G_\text{RP})$~\less~1.3~mag) to find a ZP of ($-7 \pm 3$) \muas\ when compared to the
raw GEDR3 parallaxes. The mode of the L20 correction for this sample is $-32 \pm 4$ \muas\ so, again, the L20 formula over corrects the parallaxes,
in this case by $25 \pm 5$~\muas.

\citet{StassunTorres21} continue their previous analysis using eclipsing binaries as reference objects.
\citet{Stassun18} found an offset of ($-82 \pm 33$) \muas\  based on 89 stars from GDR2 while, while their
latest analysis using GEDR3 indicates an offset of ($-37 \pm 20)$ \muas\ based on 76 objects (5~\less~$G$~\less 12~mag,
$-0.1$~\less~$(G_\text{BP} - G_\text{RP})$~\less~2.2~mag).
After applying the L20 correction the PZPO becomes ($-15 \pm 18$) \muas, indicating no over- or under-correction.

\citet{Huang21} use a sample of over 69~000 primary red clump (PRC; 9.5~\less~$G$~\less~15~mag, $1.32~\less~\nu_{\rm eff}~\less 1.5$, or about
$1.0~\less$~$(G_\text{BP} - G_\text{RP})$~$\less~2.2$~mag) stars based on LAMOST data from \citet{Huang20}.
The distances come from \citet{Schonrich19} that are based on a Bayesian analysis of DR2 data and include an PZPO of $-0.054$~mas.
The reference distance is compared to the raw GEDR3 parallax, and the GEDR3 parallax after applying the L20 correction.
The difference ({\it Gaia} - PRC) is $-26$ in the former and $+3.7$~\muas\ in the latter case (no errors are reported).
They also show the trends of the parallax difference against $G$, $\nu_{\rm eff}$, and ecliptic latitude.
They show that applying the L20 correction removes some of these trends (in particular against $G$ magnitude), but not all, and
show that there is a trend with ecliptic longitude, especially for ecliptic latitudes $<$30\degr.

The aim of the present paper is to have an independent (and alternative) investigation into the PZPO, and in particular
into the spatial dependence. This will be achieved by using a large sample of reliable QSOs (selected differently from
the sample used in various GEDR3 papers).
In addition physical binaries will be considered to derive the dependence of the PZPO on $G$ magnitude.
At the bright end the PZPO derived in the present paper will be applied to a sample of stars that has not
been systematically considered in previous works, namely stars that have an independent parallax measurement
from the {\it Hubble Space Telescope} (HST).
In addition, the PZPO will be applied to the sample of CCs by \cite{RiessGEDR3},
and results will be compared to using the correction in L20.

The paper is structured as follows.
In Section~\ref{S-Meth} the main methodology is introduced.
In Section~\ref{S-Sam} the sample of stars is described.
Section~{\ref{S-Res}} presents the results of the calculations and the derivation of the PZPO, while
 Sect.~{\ref{S-Appl}} applies the PZPO to the QSO sample itself and the two samples of stars.
A brief discussion and summary concludes the paper.

\section{Methodology}
\label{S-Meth}

The parallax zero point offset is defined through 
\begin{equation}
  \pi_{\rm t} =  \pi_{\rm o} - {\rm ZP}
  \label{Eq-basic}
\end{equation}
where  $\pi_{\rm t}$ is the true parallax and $\pi_{\rm o}$ the observed parallax (as listed in the GEDR3 catalogue).
The  PZPO is parameterised in the present paper as a sum of linear functions that are assumed to hold over a range in magnitudes:
\begin{eqnarray}
{\rm ZP} &=&  C_{\rm 0}(\alpha,\delta) + C_{\rm 1} \; (G - {\rm Gref}) + C_{\rm 2} \; (G - {\rm Gref})^{2} \nonumber \\
         & & \quad {} \quad {} \quad {} \;  + C_{\rm 3}  \; ((G_\text{BP} - G_\text{RP}) -{\rm BRref}) \\
         &=& \pi_{\rm o} - \pi_{\rm t}   \nonumber
\label{Eq-ZP2}
\end{eqnarray}
where %Gmag, Bp and Rp are the magnitudes in the G-, Bp-, and Rp-band,
Gref= 20.0~mag and BRref= 0.8~mag are reference values.
They are chosen to represent the typical colours of QSOs.

In the present analysis also binaries are considered. In that case the true parallaxes for the two components are
essentially the same and as they are essentially at the same position on the sky the $C_{\rm 0}$ term may be assumed to cancel out.
The difference of Eq.~2 %\ref{Eq-ZP2} 
for the two components (labelled as subscripts as primary, p, and secondary, s) becomes:

\begin{eqnarray}
\Delta {\rm ZP} &=&  C_{\rm 1} \; (G_{p} - G_{s})  + \nonumber \\
         & &  C_{\rm 2} \; ((G_{p} - G{\rm ref})^{2} - (G_{s} - G{\rm ref})^{2}) + \nonumber \\
         & &  C_{\rm 3}  \; ({(G_\text{BP} - G_\text{RP})_{p} - (G_\text{BP} - G_\text{RP})_{s})}) + \epsilon \\
         &=& (\pi_{\rm o})_{p} - (\pi_{\rm o})_{s}   \nonumber
\end{eqnarray}
where $\epsilon$ is a term that can be thought off as the spatial correlation on the extent of the binary separation. 

The difference between the formalism outlined in Appendix~A in L20 and that is used here is two-fold.
The main difference is that the spatial dependence is made explicit here rather than using a second-order polynomial in $\sin \beta$.
The other difference is that the spatial, magnitude and colour dependence are assumed to be separable while the L20 correction
allows for cross-terms.

The term $C_{\rm 0}$ is allowed to vary over the sky and the HEALPix formalism (\citealt{HEALPix05},
the NESTED variant) is used to transform $(\alpha, \delta)$ to a sky pixel. This is done using a implementation
in python, HEALPy \citep{HEALPy19}.
The number of sky pixels depends on the chosen resolution; resolution levels 0, 1, 2, 3, and 4 are considered in the present paper which 
correspond to 12, 48, 192, 768 and 3072 pixels, respectively. The highest resolution corresponds to a mean spacing of 3.7 degrees between sky pixels.

The fitting of Eqs.~2 or 3 to the data is done with the singular value decomposition algorithm (routine {\it  svdfit}) as
implemented in Fortran77 in \citet{Press1992}. This algorithm minimises the $\chi^2$ taking into account the errors in the ordinate
(the parallax (difference)) and gives the best-fit parameter values with error bars.
In order to also consider the errors in magnitudes and colours Monte Carlo simulations 
are performed where new datasets
are generated taking into account Gaussian errors in the parallaxes, magnitudes, and colours.
The parameter values quoted below (in Table~\ref{Tab:Res}) are the median values for the parameters among these simulations with as error
the dispersion among the parameter values,
calculated as 1.4826 times the median-absolute-deviation (MAD), equivalent to 1$\sigma$ in a Gaussian distribution.

\section{The sample} 
\label{S-Sam}

In order to apply Eq.~2 or 3 and determine ZP as a function of sky position, magnitude and colour a large sample of sources
with known true distances is required.
In this paper QSOs, physical binaries, and stars with an independent trigonometric parallax determination will be considered.

For the samples discussed below the following parameters were retrieved from the
GEDR3 main catalogue\footnote{The data is downloaded from the copy available via ViZier at the CDS.}:
parallax and parallax error ({\tt parallax, parallax\_error}), 
proper motion in Ra and Dec with errors ({\tt pmra, pmra\_error, pmdec, pmdec\_error}),
the source identifier ({\tt source\_id})\footnote{The {\tt source\_id} can also be used to determine the pixel in the HEALPix scheme, as
{\tt source\_id}/(2$^{35} \cdot 4^{(12-{\rm level})})$. \label{FN2}},
which parameters have been solved for ({\tt astrometric\_params\_solved}; five- and six-parameter solutions are relevant for
the present paper, see \citealt{GEDR3_LindegrenAS}),
the renormalised unit weight error ({\tt RUWE}), 
the goodness-of-fit (GOF, {\tt astrometric\_gof\_al}),
the effective wavenumber of the source used in the astrometric solution ($\nu_{\rm eff}$, {\tt nu\_eff\_used\_in\_astrometry}), the
astrometrically estimated pseudocolour of the source  ({\tt pseudocolour)}, the
$G$, $G_\text{BP}$, and $G_\text{RP}$ magnitudes with errors ({\tt phot\_g\_mean\_mag, phot\_g\_mean\_mag\_error,
  phot\_bp\_mean\_mag, phot\_bp\_mean\_mag\_error, phot\_rp\_mean\_mag, phot\_rp\_mean\_mag\_error}).

\subsection{Quasars}

Several of the GEDR3-team papers use a QSO sample \citep{GEDR3_Klioner,GEDR3_LindegrenAS,GEDR3_LindegrenZP} but the paper
describing the selection of this sample has not been published at the time of submission.
What is available is the list of 1.61 million {\tt source\_id}s which contains a reference to a catalogue but without any quality flag.
For this reason, and because this project started before the release of GEDR3, a different QSO sample was created.

The Million Quasars (Milliquas) catalog (version 7.0b, \citealt{Flesch19}) is used which contains of order
830~000 type-I QSOs and AGN plus about 500~000  quasar candidates.
From the full catalogue the 1.37 million objects with a confirmed redshift $>0.1$ or a probability of being a quasar of $>$98\%
are selected.
The cross-match facility (xMatch) at the Centre des Donn\'ees (CDS) in Strasbourg was used to match this list with GEDR3
using a search radius of 0.15\arcsec, and this returned 998~220 matches, of which 855~518 QSOs have a
parallax, $G$, $G_\text{BP}$, and $G_\text{RP}$ magnitude available.
The true parallax for these sources is assumed to be zero.

First the general properties of the QSO sample are discussed in particular the distribution of the
GOF and the RUWE. It is recalled that the GOF parameter should follow a Gaussian distribution with zero
mean and unit dispersion \citep{Wilson31}. In GDR2 this was not the case due to the "degree-of-freedom" bug
(see Appendix~A in \citealt{Lindegren18} and the discussion in \citealt{Gr_GDR2}).
The RUWE was introduced after GDR2 \citep{LindegrenTN} as an other quality indicator of the astrometric solution.
It compared the unit-weight-error (UWE, the square-root of the reduced $\chi^2$) to that of a sample of
unproblematic stars as a function of magnitude and colour. Although the $\chi^2$ values were actually
numerically incorrect the ratio of the UWEs was probably deemed representative of the relative quality
of the astrometric solution.
In Gaia EDR3, the "degree-of-freedom" bug has been corrected and the GOF is the main parameter to describe the
quality of the astrometric solution.

Although the QSO sample was selected to be pure it still contains non-QSOs.
As QSOs should have zero parallax and proper motion (within the error bars) the following conditions
are applied:
\begin{enumerate}
\item[] $\mid (\pi + 0.0202) \mid / \sigma_{\pi} < 5 $
\item[] $({\rm PMRA}/ \sigma_{\rm PMRA})^2 + ({\rm PMDE}/ \sigma_{\rm PMDE})^2 < 25 $
\end{enumerate}
where $-0.0202$~mas is the average offset of QSOs (see below).
Similar cuts were applied to the QSO sample in \cite{GEDR3_LindegrenZP}, but no selection on
{\tt astrometric\_params\_solved} and $\nu_{\rm eff}$ is applied.
The 5$\sigma$ limit on the parallax and proper motions (PMs) in Ra and Dec corresponds to
the expected level of outliers following Chauvenet's criterion with $\sim800~000$ objects.

\begin{figure}
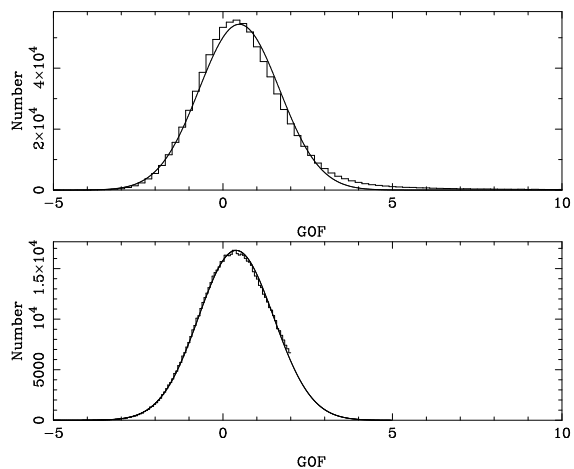

\centering

\begin{minipage}{0.4\textwidth}
\resizebox{\hsize}{!}{\includegraphics{GOF_par_PM.ps}}

\resizebox{\hsize}{!}{\includegraphics{GOF_par_PM_GOFtwo.ps}}
\end{minipage}

\caption{Distribution of the GOF with a Gaussian fit.
  Top panel, the about 841~000 QSO with parallaxes and PMs consistent with zero.
  A significant tail toward large GOF is visible.
  Lower panel, the fit with GOF restricted to $<+2.0$.
}
\label{Fig:GOF}
\end{figure}

\begin{figure}
\centering

\begin{minipage}{0.4\textwidth}
\resizebox{\hsize}{!}{\includegraphics{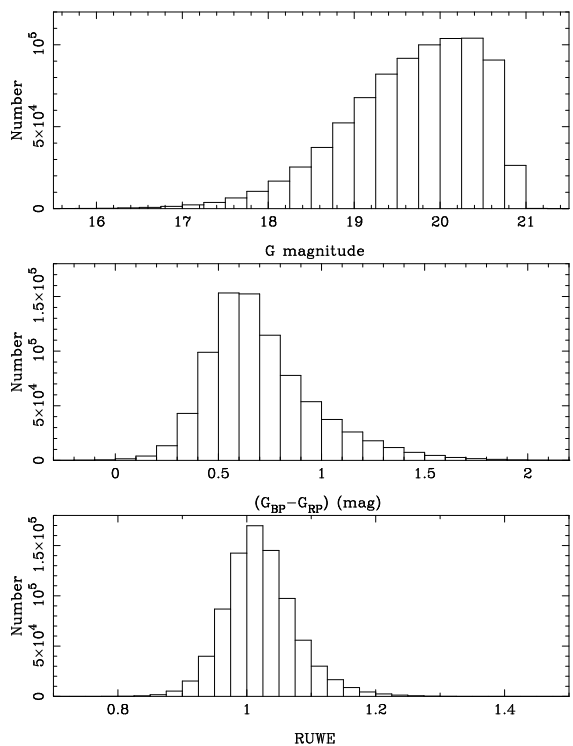}}
\end{minipage}

\caption{Distribution of $G$, ($G_\text{BP} - G_\text{RP}$) colour and RUWE for the QSO sample after applying selections on parallax,
  proper motion, GOF, and RUWE.
}
\label{Fig:QSO}
\end{figure}

The upper panel in  Figure~\ref{Fig:GOF} shows the distribution of the GOF for the 841~000 QSOs that remain after applying
the selection on parallax and PM, together with a Gaussian fit. The mean and dispersion are 0.483 and 1.187 with
negligible formal errors.
The mean is slightly larger than expected but the tail to negative GOF (the extreme value is $-5.4$) is
not inconsistent with the distribution. On the other hand the tail to larger GOF (the extreme value is +200) is obvious.
To obtain a better estimate of the mean and width of the distribution undisturbed by outliers, the lower panel
shows the distribution when the GOF and the Gaussian fit to the distribution is restricted to $< +2.0$.
The mean is 0.388 with $\sigma = 1.109$. Based on this analysis a condition $-4 <$ GOF $<5$ is imposed on all selections described
in this paper, allowing for a small excess of sources towards larger GOF.
Figure~\ref{Fig:QSO} shows the distribution in $G$, ($G_\text{BP} - G_\text{RP}$) colour, and RUWE after imposing these conditions, as well as
RUWE $<$1.4 to eliminate a few extreme outliers in that parameter.
The distribution is RUWE has a peak slightly above one, consistent with the fact that the distribution in GOF peaks slightly
above zero. It indicates that the error bars in the parallax are likely underestimated by a few percent (at least in this
range of magnitude and colour), consistent with the findings in \citet{Fabricius20} and \citet{ElBadry21}.
A final sample of 824~819 QSOs is retained. The median parallax in that sample is $-0.0202$~mas with a dispersion 
(calculated as 1.4826 times the MAD) of 0.393~mas.

\subsection{Wide binaries}
\label{SS-WB}

In this paper the catalogue of wide binaries (WBs) of \cite{ElBadry21} is used which is based on GEDR3 data.
In its raw form it contains 1.8 million candidate physical binaries.
For wide and close-by binaries the hypothesis that both components are essentially at the same physical distance may
no longer be correct. The procedure outlined in Sect.~5 of that paper is used to eliminate objects where the
true parallax difference between the two stars (Eq.~13 of that paper) is estimated to contribute more
than 5\% to the error in the observed parallax differences (eliminating about 14~000 of the 1.8 million pairs).
The sample is also restricted to the subset of about 784~000 pairs with a $<$1\% probability of chance alignment,
which is stricter than the high-confidence sample (chance alignment $<$10\%) of 1.2 million objects considered in \cite{ElBadry21}.

\cite{Fabricius20} and \citet{GEDR3_LindegrenZP} also consider samples of binaries to validate the GEDR3 results
and analyse the PZPO, respectively. Both samples are directly constructed from GEDR3 data
(but they are not identical) and careful selection is needed to reach a pure sample.

\subsection{Sources with trigonometric parallax determinations}
\label{SS-FGS}

The {\it Fine Guidance Sensor} (FGS) and the {\it Wide Field Camera 3} (WFC3) on board the {\it Hubble Space Telescope} (HST)
have been used to determine parallaxes and proper motions.
The review of \citet{Benedict17} describes the methodology and provides a list of 105 targets for which the parallax has been
determined by using the FGS. Some close binaries are included in the list with both components. One component was removed
from the list as the binary can not be resolved by \G.
Some recent works by \citet{vanBelle20} and \citet{Bond18} (on Polaris~B) are added to give a list of 102 targets.
The methodology to use the WFC3 to obtain parallaxes was described in \citet{Riess14} and \citet{Casertano16} and has been
used to obtain distances to eight CCs \citep{Riess14,Riess18}. 
Although parallaxes from \Hp\ are in general no longer competitive when compared to \G, Table~\ref{Tab-Targets} includes
one exception, Polaris~A, the most nearby CC, that had no parallax listed in GDR1 and GDR2.

The independent trigonometric parallax data and basic GEDR3 data on the 111 objects are listed in Table~\ref{Tab-Targets}.
Only one object is not listed in GEDR3 (Polaris~A), and 8 objects have no parallax listed (all these have exceptionally large GOF values).
Applying the selection on GOF and RUWE further eliminates 37 objects for a useful sample of 57 objects.

\longtab{
\footnotesize
\setlength{\tabcolsep}{1.3mm}
\begin{landscape}
\vfill
\begin{longtable}{lrcrrrrrrccc}
  \caption{\label{Tab-Targets} Stars with independent trigonometric parallaxes, ordered by Right Ascension.
    %Only a sample is shown.
  }\\
\hline
\hline
\centering
\footnotesize
Identifier & parallax  & Ref.                            & Ra                 & Dec                & Source ID         &  parallax             &  GoF     & RUWE      & $G$             & $(G_\text{BP} - G_\text{RP})$              \\
           & (mas)     &                                        & ($\deg$)           &  ($\deg$)          &                   & (mas)                 &      &          &  (mag)            &  (mag)    \\
\hline
\endfirsthead
\caption{continued.}\\
\hline\hline
\footnotesize
Identifier & parallax  & Ref.                            & Ra                 & Dec                & Source ID         &  parallax             &  GoF       & RUWE     & $G$             & $(G_\text{BP} - G_\text{RP})$   \\
           & (mas)     &                                        & ($\deg$)             &  ($\deg$)        &                   & (mas)                  &       &        &  (mag)            &  (mag)  \\
\hline
\endhead

%\multicolumn{11}{c}{Trigonometric parallaxes} \\

GJ1005A & 166.6 $\pm$ 0.3 & 1 &    3.869880  & -16.136572  &  2368293487261055488 &                  &  809.35  &      & 10.268  $\pm$  0.005  &   2.819  $\pm$  0.005   \\ 
GJ22C & 99.2 $\pm$ 0.6 & 1 &    8.142840  & +67.233278  &   527956488339113600 &   100.397  $\pm$ 0.037  &   12.87  &    1.54  & 11.107  $\pm$  0.003  &   2.620  $\pm$  0.012   \\ 
GJ22A & 99.2 $\pm$ 0.6 & 1 &    8.143031  & +67.234328  &   527956488340229632 &   101.086  $\pm$ 0.461  &   91.24  &    5.89  &  9.567  $\pm$  0.003  &   2.304  $\pm$  0.005   \\ 
$\upsilon$ Aan & 73.71 $\pm$ 0.1 & 1 &   24.198321  & +41.403762  &   348020482735930112 &    74.194  $\pm$ 0.208  &   77.72  &    7.25  &  3.966  $\pm$  0.003  &   0.723  $\pm$  0.007   \\ 
VX Per   & 0.42 $\pm$ 0.0744 &    5  &   31.952002  & +58.443534  &   506779550797525760 &     0.364  $\pm$ 0.017  &    3.49  &    1.12  &  8.894  $\pm$  0.009  &   1.551  $\pm$  0.048   \\ 
RW Tri & 2.93 $\pm$ 0.33 & 1 &   36.400769  & +28.097391  &   130692247044752768 &     3.267  $\pm$ 0.022  &    6.10  &    1.34  & 13.246  $\pm$  0.033  &   0.631  $\pm$  0.153   \\ 
Polaris B  & 6.26 $\pm$ 0.24 &  2  &   37.664816  & +89.260830  &   576402619921510144 &     7.287  $\pm$ 0.018  &    3.55  &    1.15  &  8.630  $\pm$  0.003  &   0.492  $\pm$  0.005   \\ 
Polaris A  & 7.62 $\pm$ 0.08 &  3  &     (37.967198) & (+89.264051)                     &   &                  &      &     &                 &                   \\ 
Feige 24 & 14.6 $\pm$ 0.4 & 1 &   38.782009  &  +3.732482  &  2503828498910129664 &    12.868  $\pm$ 0.036  &   11.01  &    1.58  & 12.210  $\pm$  0.003  &   0.193  $\pm$  0.006   \\ 
REJ 0317-853 & 34.38 $\pm$ 0.26 & 1 &   49.311809  & -85.540486  &  4613612951211823104 &    34.035  $\pm$ 0.029  &    0.69  &    1.02  & 14.754  $\pm$  0.004  &  -0.373  $\pm$  0.014   \\ 
LB 9802 & 33.28 $\pm$ 0.24 & 1 &   49.326169  & -85.542124  &  4613612951211823616 &    34.021  $\pm$ 0.019  &    2.49  &    1.09  & 14.079  $\pm$  0.003  &  -0.182  $\pm$  0.005   \\ 
GK Per & 2.1 $\pm$ 0.12 & 1 &   52.799999  & +43.904220  &   238540495056450048 &     2.306  $\pm$ 0.042  &   48.50  &    3.35  & 12.557  $\pm$  0.007  &   1.393  $\pm$  0.029   \\ 
$\epsilon$ Eri & 311.37 $\pm$ 0.11 & 1 &   53.228293  &  -9.458168  &  5164707970261890560 &   310.577  $\pm$ 0.136  &   31.41  &    2.72  &  3.466  $\pm$  0.003  &   1.140  $\pm$  0.011   \\ 
Cl* Melotte 22 HII 3030 & 7.41 $\pm$ 0.18 & 1 &   57.855681  & +23.889172  &   66481734354737792 &     7.371  $\pm$ 0.018  &    0.47  &    1.02  & 13.465  $\pm$  0.003  &   1.739  $\pm$  0.007   \\ 
Cl* Melotte 22 HII 3063 & 7.43 $\pm$ 0.16 & 1 &   57.874842  & +23.899036  &   66481837433947392 &     7.498  $\pm$ 0.025  &    8.04  &    1.30  & 13.028  $\pm$  0.003  &   1.571  $\pm$  0.007   \\ 
Cl* Melotte 22 HII 3179 & 7.45 $\pm$ 0.16 & 1 &   57.987027  & +23.901753  &   66471220274934272 &     7.565  $\pm$ 0.017  &    2.34  &    1.08  &  9.927  $\pm$  0.003  &   0.752  $\pm$  0.005   \\ 
XO-3 & 5.67 $\pm$ 0.14 & 1 &   65.469581  & +57.817210  &   470650560779277952 &     4.869  $\pm$ 0.026  &    9.03  &    1.25  &  9.743  $\pm$  0.003  &   0.565  $\pm$  0.005   \\ 
vA 310 & 20.13 $\pm$ 0.17 & 1 &   66.071101  & +18.002784  &  3314079508140198528 &    21.497  $\pm$ 0.015  &    0.69  &    1.02  &  9.650  $\pm$  0.003  &   1.248  $\pm$  0.005   \\ 
vA 383 & 21.53 $\pm$ 0.2 & 1 &   66.520077  & +15.041285  &  3311179340063437952 &    20.839  $\pm$ 0.028  &   25.01  &    1.89  & 11.470  $\pm$  0.003  &   1.912  $\pm$  0.005   \\ 
vA 472 & 21.7 $\pm$ 0.15 & 1 &   67.018941  & +13.867854  &  3307844864893938304 &    20.876  $\pm$ 0.021  &    0.68  &    1.02  &  8.840  $\pm$  0.003  &   0.990  $\pm$  0.005   \\ 
vA 548 & 20.69 $\pm$ 0.17 & 1 &   67.379585  & +16.244692  &  3312899491645515776 &    21.818  $\pm$ 0.021  &    6.31  &    1.24  &  9.932  $\pm$  0.003  &   1.373  $\pm$  0.005   \\ 
vA 622 & 24.11 $\pm$ 0.3 & 1 &   67.871614  & +17.718519  &  3314150048683152896 &    22.236  $\pm$ 0.022  &    5.07  &    1.18  & 11.184  $\pm$  0.003  &   1.879  $\pm$  0.005   \\ 
vA 627 & 21.74 $\pm$ 0.25 & 1 &   67.905103  & +17.709637  &  3314151251273992832 &    28.508  $\pm$ 0.581  &  488.26  &   43.27  &  9.248  $\pm$  0.003  &   1.180  $\pm$  0.005   \\ 
vA 645 & 17.46 $\pm$ 0.21 & 1 &   67.969120  & +15.499373  &  3312564037520033792 &    21.872  $\pm$ 0.016  &    2.77  &    1.10  & 10.498  $\pm$  0.003  &   1.573  $\pm$  0.005   \\ 
HD 33636 & 35.6 $\pm$ 0.2 & 1 &   77.944347  &  +4.402925  &  3238810137558836352 &    33.798  $\pm$ 0.053  &   20.84  &    1.88  &  6.865  $\pm$  0.003  &   0.750  $\pm$  0.005   \\ 
SY Aur   & 0.428 $\pm$ 0.054 &   4  &   78.163452  & +42.831780  &   201509768065410944 &     0.427  $\pm$ 0.020  &    2.50  &    1.08  &  8.798  $\pm$  0.008  &   1.376  $\pm$  0.038   \\ 
TV Col & 2.7 $\pm$ 0.11 & 1 &   82.356418  & -32.817651  &  2901783160488793728 &     1.951  $\pm$ 0.014  &    2.96  &    1.10  & 13.981  $\pm$  0.011  &   0.456  $\pm$  0.050   \\ 
GJ1081A & 65.2 $\pm$ 0.37 & 1 &   83.329996  & +44.814683  &   207910437566174592 &                  &  647.99  &     & 11.019  $\pm$  0.004  &   2.678  $\pm$  0.005   \\ 
$\beta$ Dor & 3.14 $\pm$ 0.16 & 1 &   83.406311  & -62.489769  &  4757601523650165120 &     2.931  $\pm$ 0.139  &   47.30  &    4.54  &  3.593  $\pm$  0.014  &   1.050  $\pm$  0.053   \\ 
HD 38529 & 25.11 $\pm$ 0.19 & 1 &   86.645126  &  +1.167567  &  3219847066672970368 &    23.571  $\pm$ 0.042  &    1.31  &    1.05  &  5.748  $\pm$  0.003  &   0.913  $\pm$  0.005   \\ 
SS Aur & 5.99 $\pm$ 0.33 & 1 &   93.343503  & +47.740270  &   968824328534823936 &     3.978  $\pm$ 0.028  &   15.02  &    1.69  & 14.318  $\pm$  0.024  &   0.978  $\pm$  0.098   \\ 
RT Aur & 2.4 $\pm$ 0.19 & 1 &   97.142032  & +30.492976  &  3435571660360952704 &     1.815  $\pm$ 0.122  &   66.45  &    6.43  &  5.336  $\pm$  0.021  &   0.828  $\pm$  0.086   \\ 
GJ234A & 240.98 $\pm$ 0.4 & 1 &   97.350795  &  -2.817131  &  3117120863523946368 &                  &  761.25  &      &  9.630  $\pm$  0.005  &   3.080  $\pm$  0.012   \\ 
RR Pic & 1.92 $\pm$ 0.18 & 1 &   98.900296  & -62.640097  &  5477422099543150592 &     1.996  $\pm$ 0.021  &    2.50  &    1.10  & 12.425  $\pm$  0.006  &   0.013  $\pm$  0.021   \\ 
HD 47536 & 8.71 $\pm$ 0.16 & 1 &   99.448976  & -32.339444  &  5583831735369515008 &     7.990  $\pm$ 0.053  &   -3.12  &    0.88  &  4.874  $\pm$  0.003  &   1.399  $\pm$  0.006   \\ 
G250-029A & 95.59 $\pm$ 0.28 & 1 &  103.522633  & +60.867338  &  1003752587430126464 &    91.611  $\pm$ 0.503  &  130.21  &    9.05  & 10.003  $\pm$  0.006  &   2.534  $\pm$  0.005   \\ 
G193-027A & 110.2 $\pm$ 1.1 & 1 &  105.987163  & +52.697727  &   981548637301374336 &   110.826  $\pm$ 0.695  &  234.89  &   18.90  & 11.699  $\pm$  0.003  &   3.210  $\pm$  0.005   \\ 
$\zeta$ Gem & 2.78 $\pm$ 0.18 & 1 &  106.027177  & +20.570294  &  3366754155291545344 &     3.073  $\pm$ 0.218  &   21.96  &    2.78  &  3.540  $\pm$  0.006  &   0.987  $\pm$  0.030   \\ 
SS CMa   & 0.389 $\pm$ 0.0287 &   5  &  111.529960  & -25.257297  &  5616601820448126336 &     0.286  $\pm$ 0.013  &    3.97  &    1.11  &  9.563  $\pm$  0.007  &   1.673  $\pm$  0.032   \\ 
 X Pup   & 0.277 $\pm$ 0.0469 &    5  &  113.195965  & -20.909682  &  5620098679741674496 &     0.376  $\pm$ 0.020  &    1.19  &    1.04  &  8.348  $\pm$  0.012  &   1.697  $\pm$  0.052   \\ 
YY Gem         & 67.22 $\pm$ 0.4 &  6   &  113.654977  & +31.869063  &   892348454394856064 &    66.311  $\pm$ 0.024  &    0.16  &    1.01  &  8.296  $\pm$  0.003  &   1.932  $\pm$  0.008   \\ 
U Gem & 9.96 $\pm$ 0.37 & 1 &  118.771670  & +22.001222  &   674214551557961984 &    10.705  $\pm$ 0.034  &   31.00  &    2.41  & 13.903  $\pm$  0.011  &   1.280  $\pm$  0.049   \\ 
$\rho^1$ Cnc & 79.78 $\pm$ 0.3 & 1 &  133.146761  & +28.329783  &   704967037090946688 &    79.448  $\pm$ 0.043  &   -3.81  &    0.86  &  5.733  $\pm$  0.003  &   1.009  $\pm$  0.005   \\ 
PN A66 31 & 1.61 $\pm$ 0.21 & 1 &  133.554806  &  +8.898001  &   597324024095840512 &     1.842  $\pm$ 0.055  &    0.95  &    1.03  & 15.475  $\pm$  0.003  &  -0.475  $\pm$  0.006   \\ 
VY Pyx & 6.44 $\pm$ 0.23 & 1 &  133.623514  & -23.521695  &  5653136461526964224 &     3.950  $\pm$ 0.019  &   -3.57  &    0.89  &  7.107  $\pm$  0.005  &   0.811  $\pm$  0.017   \\ 
HIP 46120 & 15.01 $\pm$ 0.12 & 1 &  141.092846  & -80.517086  &  5195968563310843008 &    14.776  $\pm$ 0.014  &   -1.60  &    0.94  &  9.938  $\pm$  0.003  &   0.814  $\pm$  0.005   \\ 
$\ell$ Car & 2.01 $\pm$ 0.2 & 1 &  146.311593  & -62.507867  &  5250032958818831360 &     1.984  $\pm$ 0.110  &   24.35  &    2.39  &  3.471  $\pm$  0.014  &   1.503  $\pm$  0.047   \\ 
HD 84937 & 12.24 $\pm$ 0.2 & 1 &  147.235464  & +13.740818  &   615943806835727872 &    13.498  $\pm$ 0.044  &    1.94  &    1.10  &  8.207  $\pm$  0.003  &   0.606  $\pm$  0.005   \\ 
VY Car   & 0.586 $\pm$ 0.0438 &    5  &  161.136160  & -57.565357  &  5351161399793209984 &     0.553  $\pm$ 0.017  &   -2.35  &    0.92  &  7.338  $\pm$  0.009  &   1.511  $\pm$  0.038   \\ 
XY Car   & 0.438 $\pm$ 0.0469 &    5  &  165.566881  & -64.262893  &  5240441472232302848 &     0.378  $\pm$ 0.014  &    2.08  &    1.07  &  8.941  $\pm$  0.009  &   1.549  $\pm$  0.043   \\ 
HIP 54639 & 11.12 $\pm$ 0.11 & 1 &  167.747457  &  +6.417567  &  3817965105665685504 &    12.141  $\pm$ 0.020  &    2.60  &    1.12  & 11.131  $\pm$  0.003  &   0.990  $\pm$  0.005   \\ 
SU Dra & 1.42 $\pm$ 0.16 & 1 &  174.485330  & +67.329393  &  1058066262817534336 &     1.332  $\pm$ 0.014  &    0.35  &    1.01  &  9.784  $\pm$  0.012  &   0.622  $\pm$  0.050   \\ 
GJ469A & 76.41 $\pm$ 0.46 & 1 &  187.237102  &  +8.424238  &  3902745286187581312 &    72.266  $\pm$ 0.696  &  201.12  &   21.31  & 10.866  $\pm$  0.006  &   2.781  $\pm$  0.006   \\ 
AM CVn & 1.65 $\pm$ 0.3 & 1 &  188.727770  & +37.628978  &  1519860699806445184 &     3.311  $\pm$ 0.030  &    1.86  &    1.07  & 14.059  $\pm$  0.003  &  -0.283  $\pm$  0.009   \\ 
EX Hya & 15.5 $\pm$ 0.29 & 1 &  193.100317  & -29.248754  &  6185040879503491584 &    17.572  $\pm$ 0.017  &    2.65  &    1.08  & 13.246  $\pm$  0.008  &   0.416  $\pm$  0.034   \\ 
GP Com & 13.34 $\pm$ 0.33 & 1 &  196.425059  & +18.017867  &  3938156295111047680 &    13.731  $\pm$ 0.045  &    1.21  &    1.04  & 15.929  $\pm$  0.004  &   0.021  $\pm$  0.014   \\ 
V803 Cen & 2.88 $\pm$ 0.24 & 1 &  200.935556  & -41.741460  &  6137049739573759872 &     3.489  $\pm$ 0.060  &   12.95  &    1.66  & 15.731  $\pm$  0.106  &   0.232  $\pm$  0.475   \\ 
NSVS 0103       & 14.92 $\pm$ 0.53 &  6   &  206.392715  & +79.397010  &  1715299716278321408 &    16.572  $\pm$ 0.018  &   11.67  &    1.51  & 12.363  $\pm$  0.004  &   2.451  $\pm$  0.012   \\ 
NSVS 0103-REF68  & 14.84 $\pm$ 0.66 &  6   &  206.476952  & +79.387440  &  1715287999607537408 &    16.576  $\pm$ 0.018  &    4.20  &    1.17  & 14.486  $\pm$  0.003  &   2.741  $\pm$  0.007   \\ 
CR Boo & 2.97 $\pm$ 0.34 & 1 &  207.229937  &  +7.959982  &  3721961488404743040 &     2.844  $\pm$ 0.037  &   -1.98  &    0.93  & 15.467  $\pm$  0.052  &   0.066  $\pm$  0.208   \\ 
Proxima Cen & 768.7 $\pm$ 0.3 & 1 &  217.392321  & -62.676075  &  5853498713190525696 &   768.067  $\pm$ 0.050  &   -1.20  &    0.97  &  8.985  $\pm$  0.003  &   3.805  $\pm$  0.006   \\ 
G166-037 & 5.2 $\pm$ 0.7 & 1 &  218.712738  & +25.166043  &  1255095276181144320 &     5.155  $\pm$ 0.014  &    1.59  &    1.06  & 12.453  $\pm$  0.003  &   0.911  $\pm$  0.005   \\ 
HD 128311 & 60.53 $\pm$ 0.15 & 1 &  219.003258  &  +9.745402  &  1176209886733406592 &    61.279  $\pm$ 0.043  &    5.14  &    1.31  &  7.181  $\pm$  0.003  &   1.164  $\pm$  0.005   \\ 
HD 132475 & 10.18 $\pm$ 0.21 & 1 &  224.954678  & -22.014952  &  6232043867720079616 &    10.671  $\pm$ 0.025  &    3.22  &    1.16  &  8.391  $\pm$  0.003  &   0.794  $\pm$  0.005   \\ 
HD 136118 & 19.12 $\pm$ 0.22 & 1 &  229.730587  &  -1.592288  &  4415515934099120768 &    19.812  $\pm$ 0.034  &    8.15  &    1.43  &  6.813  $\pm$  0.003  &   0.690  $\pm$  0.005   \\ 
GU Boo         & 3.15 $\pm$ 0.56 &  6   &  230.478556  & +33.935715  &  1278589709364139520 &     6.187  $\pm$ 0.011  &    1.39  &    1.04  & 13.044  $\pm$  0.003  &   1.663  $\pm$  0.009   \\ 
HP Lib & 5.07 $\pm$ 0.33 & 1 &  233.971000  & -14.220117  &  6265476408553544320 &     3.567  $\pm$ 0.031  &    2.31  &    1.11  & 13.603  $\pm$  0.003  &  -0.153  $\pm$  0.006   \\ 
HD 140283 & 17.15 $\pm$ 0.14 & 1 &  235.757857  & -10.934848  &  6268770373590148224 &    16.267  $\pm$ 0.026  &    1.31  &    1.06  &  7.036  $\pm$  0.003  &   0.759  $\pm$  0.005   \\ 
G16-025 & 3.8 $\pm$ 1 & 1 &  240.339103  &  +5.393949  &  4425854676297423104 &     3.433  $\pm$ 0.014  &    0.75  &    1.03  & 13.146  $\pm$  0.003  &   0.849  $\pm$  0.005   \\ 
GJ623A & 125 $\pm$ 0.3 & 1 &  246.046510  & +48.350869  &  1411178510887026048 &                  &  775.27  &      &  9.254  $\pm$  0.003  &   2.401  $\pm$  0.007   \\ 
UV Oct & 1.71 $\pm$ 0.1 & 1 &  248.103953  & -83.903451  &  5768557209320424320 &     1.838  $\pm$ 0.012  &    0.00  &    1.00  &  9.536  $\pm$  0.010  &   0.706  $\pm$  0.043   \\ 
CM Dra        & 68.23 $\pm$ 0.38 &  6   &  248.575581  & +57.167574  &  1431176943768690816 &    67.288  $\pm$ 0.034  &    9.09  &    1.42  & 11.491  $\pm$  0.003  &   2.924  $\pm$  0.005   \\ 
CM Dra-REF47  & 65.1 $\pm$ 1.4 &  6   &  248.580692  & +57.174461  &  1431176943768691328 &    67.354  $\pm$ 0.021  &    0.36  &    1.01  & 14.849  $\pm$  0.003  &   0.685  $\pm$  0.005   \\ 
TRES HER0      & 5.58 $\pm$ 0.53 &  6   &  252.586282  & +46.650513  &  1407718450873494784 &     7.158  $\pm$ 0.017  &    2.42  &    1.09  & 14.547  $\pm$  0.003  &   2.585  $\pm$  0.008   \\ 
HIP 87062 & 8.21 $\pm$ 0.11 & 1 &  266.867655  &  -8.781545  &  4165370682239910144 &     8.718  $\pm$ 0.019  &   -3.96  &    0.88  & 10.352  $\pm$  0.003  &   0.889  $\pm$  0.005   \\ 
X Sgr & 3 $\pm$ 0.18 & 1 &  266.890075  & -27.830835  &  4057701830728920064 &     2.806  $\pm$ 0.140  &    5.11  &    1.22  &  4.327  $\pm$  0.008  &   1.098  $\pm$  0.029   \\ 
HIP 87788 & 10.83 $\pm$ 0.13 & 1 &  268.993631  & -16.411888  &  4144902306908889600 &    10.760  $\pm$ 0.016  &   -7.45  &    0.70  & 11.095  $\pm$  0.003  &   0.915  $\pm$  0.005   \\ 
Barnard Star & 545.4 $\pm$ 0.3 & 1 &  269.448503  &  +4.739420  &  4472832130942575872 &   546.976  $\pm$ 0.040  &    1.89  &    1.08  &  8.194  $\pm$  0.003  &   2.834  $\pm$  0.005   \\ 
W Sgr & 2.28 $\pm$ 0.2 & 1 &  271.255132  & -29.580110  &  4050309195613114624 &     2.365  $\pm$ 0.176  &   35.76  &    3.95  &  4.585  $\pm$  0.019  &   1.108  $\pm$  0.078   \\ 
DQ Her & 2.59 $\pm$ 0.21 & 1 &  271.876040  & +45.859101  &  2116226254706461568 &     2.016  $\pm$ 0.017  &    1.50  &    1.06  & 14.589  $\pm$  0.011  &   0.462  $\pm$  0.055   \\ 
WZ Sgr   & 0.512 $\pm$ 0.0373 &    5  &  274.248822  & -19.075831  &  4094784475310672128 &     0.574  $\pm$ 0.028  &   -1.14  &    0.94  &  7.682  $\pm$  0.016  &   1.779  $\pm$  0.065   \\ 
Y Sgr & 2.13 $\pm$ 0.29 & 1 &  275.345761  & -18.860034  &  4096107909387492992 &     1.975  $\pm$ 0.058  &   13.06  &    1.76  &  5.475  $\pm$  0.012  &   1.166  $\pm$  0.050   \\ 
V603 Aql & 4.01 $\pm$ 0.14 & 1 &  282.227705  &  +0.584085  &  4266547566124966912 &     3.106  $\pm$ 0.035  &    0.31  &    1.01  & 11.867  $\pm$  0.013  &   0.172  $\pm$  0.060   \\ 
V1223 Sgr & 1.96 $\pm$ 0.18 & 1 &  283.759644  & -31.163883  &  6760253239457454592 &     1.745  $\pm$ 0.024  &    1.66  &    1.08  & 13.029  $\pm$  0.006  &   0.135  $\pm$  0.026   \\ 
$\kappa$ Pav & 5.57 $\pm$ 0.28 & 1 &  284.237527  & -67.233423  &  6434564460631076864 &     5.245  $\pm$ 0.122  &   37.63  &    2.29  &  4.263  $\pm$  0.010  &   0.915  $\pm$  0.041   \\ 
FF Aql & 2.81 $\pm$ 0.18 & 1 &  284.561446  & +17.360872  &  4514145288240593408 &     1.906  $\pm$ 0.071  &    1.51  &    1.05  &  5.171  $\pm$  0.005  &   1.056  $\pm$  0.018   \\ 
GJ748A & 98.4 $\pm$ 0.3 & 1 &  288.068738  &  +2.884048  &  4268226078065241600 &                  & 1152.54  &      &  9.886  $\pm$  0.010  &   2.624  $\pm$  0.014   \\ 
RR Lyr & 3.77 $\pm$ 0.13 & 1 &  291.365640  & +42.783489  &  2125982599343482624 &     3.985  $\pm$ 0.026  &    0.93  &    1.04  &  7.619  $\pm$  0.015  &   0.574  $\pm$  0.062   \\ 
XZ Cyg & 1.67 $\pm$ 0.17 & 1 &  293.122787  & +56.388085  &  2142052889490819328 &     1.586  $\pm$ 0.015  &    4.09  &    1.17  &  9.914  $\pm$  0.006  &   0.639  $\pm$  0.031   \\ 
 S Vul   & 0.322 $\pm$ 0.0396 &    5  &  297.099176  & +27.286481  &  2027971514401523456 &     0.205  $\pm$ 0.020  &    1.13  &    1.03  &  8.152  $\pm$  0.007  &   2.231  $\pm$  0.033   \\ 
GJ1245A & 219.9 $\pm$ 0.5 & 1 &  298.479759  & +44.412330  &  2079074130463898624 &                  & 1646.33  &      & 11.535  $\pm$  0.003  &   3.701  $\pm$  0.007   \\ 
GJ1245C & 219.9 $\pm$ 0.5 & 1 &  298.481926  & +44.412906  &  2079073928612821760 &   214.575  $\pm$ 0.048  &   15.35  &    1.61  & 11.908  $\pm$  0.003  &   3.823  $\pm$  0.005   \\ 
NGC6853 & 2.47 $\pm$ 0.16 & 1 &  299.901564  & +22.721214  &  1827256624493300096 &     2.570  $\pm$ 0.037  &    5.53  &    1.16  & 14.037  $\pm$  0.003  &  -0.541  $\pm$  0.005   \\ 
HIP 98492 & 3.49 $\pm$ 0.14 & 1 &  300.139629  &  +9.352701  &  4299974407538484096 &     2.660  $\pm$ 0.018  &   -4.13  &    0.85  & 11.373  $\pm$  0.003  &   0.898  $\pm$  0.005   \\ 
WZ Sge & 22.97 $\pm$ 0.15 & 1 &  301.902433  & +17.703984  &  1809844934461976832 &    22.104  $\pm$ 0.030  &    1.29  &    1.04  & 15.181  $\pm$  0.004  &   0.170  $\pm$  0.017   \\ 
GJ791.2A & 113.4 $\pm$ 0.2 & 1 &  307.454338  &  +9.689504  &  1752805741531173632 &                  & 1078.85  &     & 11.485  $\pm$  0.003  &   3.168  $\pm$  0.006   \\ 
T Vul & 1.9 $\pm$ 0.23 & 1 &  312.867677  & +28.250482  &  1857884212378132096 &     1.688  $\pm$ 0.058  &    5.28  &    1.20  &  5.500  $\pm$  0.010  &   0.824  $\pm$  0.041   \\ 
HIP 103269 & 14.12 $\pm$ 0.1 & 1 &  313.820146  & +42.298456  &  2065901676227318272 &    13.960  $\pm$ 0.012  &   -1.26  &    0.95  & 10.091  $\pm$  0.003  &   0.844  $\pm$  0.005   \\ 
HD 202206 & 22.98 $\pm$ 0.13 & 1 &  318.740516  & -20.789745  &  6832155218215202944 &    21.939  $\pm$ 0.028  &    0.82  &    1.03  &  7.922  $\pm$  0.003  &   0.848  $\pm$  0.005   \\ 
GJ831A & 125.3 $\pm$ 0.3 & 1 &  322.832981  &  -9.790965  &  6894054664842632448 &                  & 1186.79  &    & 10.471  $\pm$  0.003  &   3.083  $\pm$  0.008   \\ 
HIP 106924 & 14.47 $\pm$ 0.1 & 1 &  324.813858  & +60.284865  &  2203746967971153024 &    15.019  $\pm$ 0.012  &   -5.10  &    0.79  & 10.155  $\pm$  0.003  &   0.871  $\pm$  0.005   \\ 
SS Cyg & 8.3 $\pm$ 0.41 & 1 &  325.679037  & +43.586222  &  1972957892448494592 &     8.854  $\pm$ 0.030  &   33.77  &    2.33  & 11.671  $\pm$  0.014  &   1.177  $\pm$  0.078   \\ 
HIP 108200 & 12.4 $\pm$ 0.09 & 1 &  328.821293  & +32.645318  &  1946297900868982016 &    12.373  $\pm$ 0.014  &    3.43  &    1.09  & 10.783  $\pm$  0.003  &   0.932  $\pm$  0.005   \\ 
RU Peg & 3.55 $\pm$ 0.26 & 1 &  333.510604  & +12.703148  &  2727974767550030080 &     3.662  $\pm$ 0.021  &    5.72  &    1.29  & 12.334  $\pm$  0.012  &   0.964  $\pm$  0.058   \\ 
PN DeHt5 & 2.9 $\pm$ 0.15 & 1 &  334.890158  & +70.934134  &  2229624931896924160 &     2.982  $\pm$ 0.036  &    5.46  &    1.22  & 15.462  $\pm$  0.003  &  -0.304  $\pm$  0.006   \\ 
HD 213307 & 3.65 $\pm$ 0.15 & 1 &  337.288645  & +58.404113  &  2200153214212849024 &     3.454  $\pm$ 0.051  &    9.58  &    1.42  &  6.300  $\pm$  0.003  &  -0.025  $\pm$  0.005   \\ 
$\delta$ Cep & 3.66 $\pm$ 0.15 & 1 &  337.292885  & +58.415208  &  2200153454733285248 &     3.555  $\pm$ 0.147  &   31.03  &    2.71  &  3.851  $\pm$  0.014  &   0.971  $\pm$  0.057   \\ 
NGC7293 & 4.64 $\pm$ 0.27 & 1 &  337.410790  & -20.837167  &  6628874205642084224 &     5.012  $\pm$ 0.044  &   -1.02  &    0.95  & 13.459  $\pm$  0.003  &  -0.588  $\pm$  0.006   \\ 
RZ Cep & 2.12 $\pm$ 0.16 & 1 &  339.805873  & +64.859351  &  2211629018927324288 &     2.401  $\pm$ 0.012  &   -0.29  &    0.99  &  9.294  $\pm$  0.009  &   0.740  $\pm$  0.039   \\ 
GJ 876 & 214.6 $\pm$ 0.2 & 1 &  343.324111  & -14.266689  &  2603090003484152064 &   214.038  $\pm$ 0.036  &    6.49  &    1.34  &  8.875  $\pm$  0.003  &   2.809  $\pm$  0.005   \\ 
$\gamma$ Cep & 74.27 $\pm$ 0.12 & 1 &  354.835781  & +77.633125  &  2281778105594488192 &    72.517  $\pm$ 0.147  &   37.24  &    3.21  &  2.943  $\pm$  0.003  &   1.257  $\pm$  0.022   \\ 

\hline
\end{longtable}
\tablefoot{
Column~1. Identifier.
Column~2. Trigonometric parallax with error.
Column~3. References for the parallax.
1=\citet{Benedict17},
2=\citet{Bond18},
3=\citet{Gr_GDR2},   
4=\citet{Riess14},
5=\citet{Riess18}, and
6=\citet{vanBelle20},
Column~4,5. Ra and Declination from GEDR3. Stars not in GEDR3 have coordinates listed between parentheses.
Column~6. source identifier from GEDR3.
Column~7. parallax with error from GEDR3.
Column~8,9. Goodness-of-fit and Renormalised Unit Weight Error.
Column~10,11. $G$ magnitude and $(G_\text{BP} - G_\text{RP})$ colour.

} 
\end{landscape}
}

%%%%%%%%%%%%%%%%%%%%%%%%%%%%%%%%%%%%%%%%%%%%%%%%%%%%%%%%%

\section{Results}
\label{S-Res}

In this section the results are being discussed related to the derivation of the PZPO correction.

\subsection{QSOs}

\begin{figure}
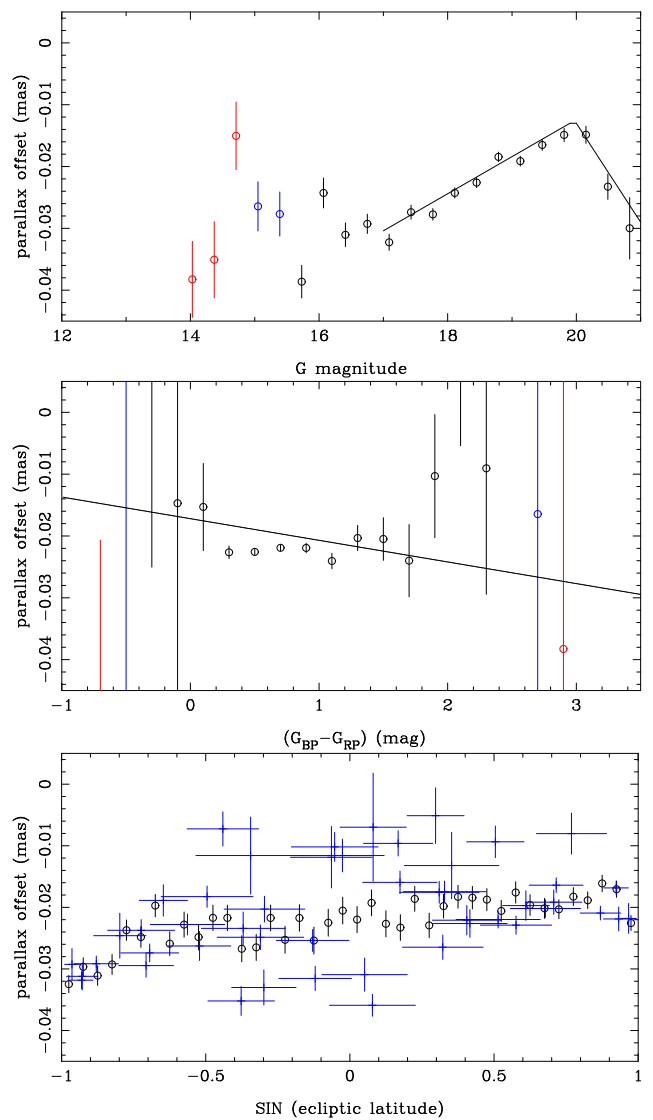

\centering
\begin{minipage}{0.45\textwidth}
\resizebox{\hsize}{!}{\includegraphics{ParOff_Gmag.ps}}
\end{minipage}
\begin{minipage}{0.45\textwidth}
\resizebox{\hsize}{!}{\includegraphics{ParOff_BpRp.ps}}
\end{minipage}
\begin{minipage}{0.45\textwidth}
\resizebox{\hsize}{!}{\includegraphics{ParOff_ELAT.ps}}
\end{minipage}

\caption{PZPO for the QSO sample as a function of $G$, $(G_\text{BP} - G_\text{RP})$ colour and ecliptic latitude (open circles).
  Only bins with $\ge$5 objects are plotted. Bins with 100 objects or less are plotted in blue, with 30 objects or less in red.
  The lines in the top panel and the lines in the middle panel are {\it not} a fit to the data, but are based on Eq.~\ref{Eq-cor}.
  The blue points in the lower panel indicate the PZPO for the 48 HEALPix level 1 pixels.
  The horizontal bar gives the range in $\sin \beta$ for each HEALPix pixel.
}
\label{Fig:QSOparoff}
\end{figure}

Figure~\ref{Fig:QSOparoff} gives another representation of the QSO sample, similar to Fig.~5 in L20.
It shows a binned version of the PZPO (weighted mean and error) as a function of $G$ magnitude,
$(G_\text{BP} - G_\text{RP})$ colour (L20 shows it as a function of $\nu_{\rm eff}$), and $\beta$.
The lines in the top and middle panels are not fits to the data but represent the finally applied corrections based on a
full analysis (see later in this section, Eqs.~\ref{Eq-cor}, \ref{Eq-cor1}).
  In the middle panel, the line does not seem to fit the points very well.
This is related to the fact that in the QSO sample the brighter QSOs are bluer than fainter ones.
In addition, the error bar on this slope is quite large.

Although the QSO sample is different from that used in GEDR3-team papers the behaviour is very similar to that shown in L20, as expected.
There is a quite noticeable correlation with $G$ for $G$ \more 17~mag, a small (if any) correlation with colour (or $\nu_{\rm eff}$), and a
relatively modest correlation with $\beta$. In particular the latter correlation is interesting.
An identical binning is used as in L20 (40 bins) and the distribution of the black open squares is quasi identical to that shown in L20.
The blue points show another representation, and the main reason why a different spatial dependence of the PZPO is
proposed here. The binning is now done based on HEALPix level 1, which has 48 pixels, very similar to the binning in L20.
The point and the vertical error bar have the same meaning,
while the horizontal line represents the range in $\beta$ for that HEALPix pixel. One can clearly observe a significantly larger spread even for
pixels with very similar ecliptic latitudes. This indicates that the PZPO is a more complicated
function than of $\beta$ alone. The result is qualitatively similar to \citet{Huang21} who demonstrate that there is a trend of the
PZPO with  ecliptic longitude, especially for $\beta <$30\degr. 

To investigate this further Eq.~2 is fit to this sample, solving for $C_0$ (only a spatial component), $C_0$+$C_1$+$C_2$, and
all parameters. This is done for several HEALPix levels, and the results are summarised in Table~\ref{Tab:Res}.
As the fitting routine is based on minimising the  $\chi^2$ one expects the value for $C_0$ to be equal to the weighted mean
of the parallaxes of all QSOs in that pixel when only the spatial component is solved for. As a sanity check to the implementation of the
numerical code it was verified that this is indeed the case.

As the distribution of known quasars is not uniform over the sky (typically underrepresented in the Galactic plane) the number of
objects per sky pixel varies strongly.
At HEALPix level 3 there  are 10 pixels with no QSOs, and 63 with 40 or less objects.
On the other hand 50\% of pixels have 810 objects or more, with a maximum of 3762.
Inspecting the error in the parallax offset and the signal-to-noise suggests that forty objects or
more are required for the results to be robust. 
The median offset over these 706 pixels is $-21.0$~\muas\ with a dispersion over the pixels (calculated via the MAD) of 12.4~\muas\
when fitting only the spatial component. Averaging only over the pixels with 100 objects or more changes the parameter and the error
by $\sim0.3$~\muas.

The first entries in Table~\ref{Tab:Res} (models 1-9) include all QSOs but based on the trends seen in Fig.~\ref{Fig:QSOparoff} models
restricted in $G$ magnitude are more realistic, and several ranges have been explored. Models 10-28 show the main the results, and based
on these the following linear correction of the parallax is proposed (in \muas/mag) in the range $G >$17~mag
($\Delta \pi$ to be subtracted from the catalogued GEDR3 parallax):

\begin{equation}
\Delta\pi=\begin{cases}
$\phantom{0}$+6.0 \;(G - 19.9)           & 17.0 \le G < 19.9\\
$\phantom{0}$+0.0                        & 19.9 \le G < 20.0\\
-16.0             \;(G - $G\text{ref}$)  & 20.0 \le G < 22.5\\
\end{cases}
\label{Eq-cor}
\end{equation}

The presence of a colour dependence is less clear. As can be seen from the results in  Table~\ref{Tab:Res} the term is
not very significant (at the 2$\sigma$ level at best).
Nevertheless, the colour correction (in \muas/mag) that will be tested is:
\begin{equation}
\Delta \pi = -3.5 \; ((G_\text{BP} - G_\text{RP}) - {\rm BRref})  \\
\label{Eq-cor1}
\end{equation}

Figure~\ref{Fig:QSOparoff} shows no real trend with magnitude for brighter magnitudes. Fitting a constant at HEALPix level 0, as
there are only $\sim$3300 QSOs brighter than 17~mag, gives a value of about $-31$~\muas\ (model 29).

Models 30-34 and 35-39 give the results when the parallaxes are corrected according to Eq.~\ref{Eq-cor}, respectively,
Eq.~\ref{Eq-cor} and \ref{Eq-cor1}. The results are listed for several HEALPix levels.
The average spatial correction of the PZPO (at $G$ = 20) is essentially independent of the chosen HEALPix level and suggests
systematic errors of \less~0.5~\muas.
Tests % at HEALPix level2
using a slope of $-6.1$, respectively, $17.0$, at the bright and faint magnitude end, and shifting
the nodes by 0.1~mag indicate global differences of \less~0.2~\muas\ and changes in the  spatial PZPO in individual
pixels of \less~0.2$\sigma$.
Adding the colour correction increases the dispersion over the pixels, suggesting again that the colour term
is not a significant factor.

The detailed results of models 30-39 are available through the CDS, and an example of the content is given in Tab.~\ref{Tab:H768p3}.  %*** zenodo *** ??
These files list the PZPO and error for each individual HEALPix pixel for levels 0, 1, 2, 3, and 4, and the number
of QSOs in each pixel.

Some properties are summarised in Tab.~\ref{Tab:Sum}.
It lists the HEALPix level, the number of pixels, the number of pixels with only 0 or 1 object, and in column~4 the range in the PZPO errors for the pixels
with 40 or more objects, which typically increases with increasing spatial resolution.

\begin{figure}
\centering

\begin{minipage}{0.46\textwidth}

\resizebox{\hsize}{!}{\includegraphics{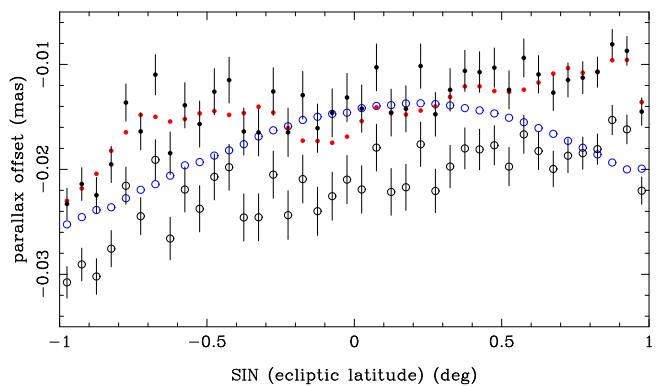}}
  
\end{minipage}

\caption{PZPO for the QSO sample with $G >$17~mag as a function of $\sin \beta$.
  Black open circles represent the observed data  (the weighted mean) to be compared to the
  blue open circles that represents the L20 correction  (the unweighted mean as the L20 correction carries no error).
  The black filled circles represent the corrected parallax data (according to Eq.~\ref{Eq-cor}) to be compared to
  the red filled circles that represent the spatial correction at $G$ = 20~mag at HEALPix level 2
  (both are weighted means, but the error in the red filled circles is too small to be visible).
}
\label{Fig:Correction}
\end{figure}

Figure~\ref{Fig:Correction} illustrates how this spatial and magnitude correction works
for the QSO sample with $G >$17~mag.
The black open circles give the observed parallaxes of the 821~000 QSOs averaged and binned over $\sin \beta$.
The blue open circles give the L20 correction for the individual QSOs again averaged and binned over $\sin \beta$, while
the black filled circles give the corrected parallaxes according to Eq.~\ref{Eq-cor} and
the red filled circles the spatial correction at HEALPix level 2.
Weighted averages are used except for the blue circles of the L20 correction. The unweighted mean is shown as the L20 correction carries no error.
%
%\sout{A first remark is that the L20 correction  appears not to be a good fit to the raw data.}
The shape of the L20 correction is due to the fact that L20 uses a second order polynominal in $\sin \beta$ (Eq.~A4 in L20 and the
discussion in their Sect.~4.1).
The spatial correction at HEALPix level 2 gives a good description of the parallaxes corrected
according to Eq.~\ref{Eq-cor} even though the fitting was done according to the HEALPix level and not specifically to ecliptic latitude.
Section~\ref{SS-QSO} discusses the results when the correction in L20 and the current one are applied to the QSO sample.

\begin{table*} 
\caption{Result of the fitting to the QSO sample.}

\begin{tabular}{cccccccc} \hline \hline 
Model &  HEALPix  &  $\langle C_0 \rangle$  & N & $C_1$       &  $C_2$          &  $C_3$    \\ %  & BIC \\
      &  level    & (\muas)   &   & (\muas/mag) & (\muas/mag$^2$) & (\muas/mag) \\ % &     \\
 \hline

 & \multicolumn{6}{c}{all $G$ magnitudes} \\

1&  2 & $-20.6 \pm 9.0$ & 190/192 & - & - & -   \\ %& 946780.062   \\
2&  2 & $-14.5 \pm 9.2$ & 190/192 & $4.00 \pm 0.25$ & - & -   \\ %& 944818.188   \\
3&  2 & $-13.8 \pm 8.9$ & 190/192 & $4.91 \pm 0.63$ & $0.24 \pm 0.13$ & -   \\ %& 944206.312   \\
4&  2 & $-14.1 \pm 9.0$ & 190/192 & $5.16 \pm 0.60$ & $0.29 \pm 0.14$ & $-3.08 \pm 1.16$ \\ % &  943326.188 \\
\\
5&  3 &  $-21.0 \pm 12.4$ & 706/768 & - & - & - \\ %& 950605.688 \\ % fort.81 : BIC 938590.188
6&  3 &  $-14.9 \pm 12.2$ & 706/768 & $3.97 \pm 0.23$ & - & - \\ %& 949230.062 \\
7&  3 &  $-14.2 \pm 12.7$ & 706/768 & $5.07 \pm 0.60$ & $0.29 \pm 0.13$ & - \\ %& 950502.562  \\
8&  3 &  $-14.9 \pm 12.5$ & 706/768 & $5.18 \pm 0.65$ & $0.28 \pm 0.13$ & $-3.61 \pm  1.23$ \\ %& 950428.188  \\
  \\

%  & \multicolumn{5}{c}{full  sample} \\                                                                                              % DeltaG= 6-20=-14
9& 4 &  $-15.1 \pm 19.0$ & 2600/3072 & $4.85 \pm 0.45$ & $0.22 \pm 0.08$ & $-3.96 \pm  1.67$ \\ %&  981842.938 \\ %Nsim=7  128 empty pixels   -24.8
\\

& \multicolumn{6}{c}{ $17 < G < 20 $ } \\

10 &1 &  $-12.1 \pm 7.4$ &  48/48  &  $5.97 \pm 0.41$ & - & - \\ %&                                172207.250 \\   

11 &2 &  $-20.7 \pm 9.0$ & 190/192 & - & - & -  \\ % &                                               174148.531 \\ %  496510 of  856026
12 &2 &  $-12.4 \pm 9.1$ & 190/192 &  $6.03 \pm 0.32$ & - & - \\ %&                                 173909.625 \\   
13 &2 &  $-12.5 \pm 9.2$ & 190/192 &  $6.11 \pm 0.44$ & - & $-3.55 \pm 1.49$ \\ %&                   174143.469 \\   
14 &2 &  $-13.4 \pm 9.2$ & 190/192 &  $4.13 \pm 1.51$ & $-0.71 \pm 0.52$ & - \\ %&                 174011.219 \\   
15 &2 &  $-13.2 \pm 9.3$ & 190/192 &  $4.54 \pm 1.71$ & $-0.53 \pm 0.54$ & $-3.17 \pm 1.35$ \\ %&  173950.969 \\   

16 &3 &  $-12.5 \pm 12.6$ & 698/768 &  $6.05 \pm 0.45$ & - & - \\ %&                                180690.531 \\   
17 &3 &  $-12.9 \pm 12.5$ & 698/768 &  $4.70 \pm 1.32$ & $-0.53 \pm 0.50$ & $-4.06 \pm 1.55$ \\ %&  180683.844 \\   
\\

%& \multicolumn{5}{c}{ $17 < Gmag < 19.5 $ } \\
%2 &  $-12.1 \pm  9.7$ & 189/192 &  $6.11 \pm 0.38$ & - & - &                                -39323.76 \\  % 304853 
%2 &  $-13.3 \pm 10.5$ & 189/192 &  $3.83 \pm 2.44$ & $-0.70 \pm 0.77$ & $-3.76 \pm 1.40$ &  -39222.62 \\   
%3 &  $-12.2 \pm 12.8$ & 683/768 &  $6.03 \pm 0.47$ & - & - &                                -32831.453 \\  % 304853 

& \multicolumn{6}{c}{ $19.875 < G < 20.125 $ } \\
18 &0 &  $-13.4 \pm  6.5$ &  12/12  & - & - & - \\ %&                                              154871.75  \\ % 
19 &1 &  $-14.3 \pm 11.3$ &  47/48  & - & - & - \\ %&                                              155262.719  \\ % nsim=1001
20 &2 &  $-12.8 \pm 22.0$ & 180/192 & - & - & - \\ %&                                              156756.500  \\ % 
21 &3 &  $-13.6 \pm 44.2$ & 595/768 &  - & - & - \\ %&                                             162741.188 \\
22 &4 &  $-13.5 \pm 62.8$ & 921/3072 &  - & - & - \\ %&                                            186949.594 \\ %Nsim=11  7h45 real time

\\
& \multicolumn{6}{c}{ $20.125 < G < 22.5 $ } \\
23 &1 &  $-22.5  \pm 23.8$ &  47/48  &  - & - & - \\ %&                                                 692702.12  \\ % nsim=1001
24 &1 &  $-16.8  \pm 22.1$ &  47/48  &  $-16.35 \pm 7.90$  & - & - \\ %&                               692646.375  \\ % nsim=1001
25 &2 &  $-13.86 \pm 33.3$ & 182/192 &  $-15.64 \pm 9.59$  & - & - \\ % &                                694229.000 \\  % 273269
26 &2 &  $-13.39 \pm 33.1$ & 182/192 &  $-16.30 \pm 7.30$  & - & $-3.71 \pm 4.17$ \\ % &                 694296.625 \\  % 273269
27 &2 &  $-13.99 \pm 34.2$ & 182/192 &  $-15.36 \pm 30.71$ & $-1.96 \pm 33.7$ & $-3.21 \pm 5.51$ \\ % & 694430.125 \\   
28 &3 &  $-15.4  \pm 51.9$ & 640/768 &  $-15.43 \pm 7.66$  & - & - \\ % &                                701149.250  \\  % 273269

\\
& \multicolumn{6}{c}{ $0 < G < 17 $ } \\
29 &0 &  $-30.6 \pm 5.5$ & 12/12 &  -                & -                & -             \\ %   &  -9039.490 \\ %  3294 QSO of  856026   NSIM=5001
\\

& \multicolumn{6}{c}{ $17 < G < 22.5$, magnitude corrected according to Eq.~\ref{Eq-cor} } \\ %FINAL -6.0 +16.0 Gmag
30 &0 &  $-11.8 \pm 3.0$ &  12/12   & - & - & - \\ % &                                             951225.500  \\ % nsim=5001   
31 &1 &  $-13.1 \pm 7.0$ &  48/48  & - & - & - \\ % &                                              951535.188  \\ % nsim=1001   82
%2 &  $-12.1 \pm 8.9$ & 190/192 & - & - & - &                                              953102.562  \\ % nsim= 101  118min
32 &2 &  $-12.8 \pm 8.9$ & 190/192 & - & - & - \\ % &                                              953285.188  \\ % nsim=1001 1167
33 &3 &  $-12.4 \pm 12.3$ & 705/768 & - & - & - \\ % &                                             960330.062  \\ % nsim= 101 1952
%&4 &  $-12.7 \pm 19.7$ & 2599/3072 & - & - & - &                                           989045.812  \\ % nsim=  11 3539m=2.5d
34 &4 &  $-13.1 \pm 20.1$ & 2599/3072 & - & - & - \\ % &                                           989780.0622  \\ % nsim=  15  4766m44.

%&2 &  $-12.6 \pm 9.1$ & 190/192 & - & - & - &                                              953327.688 \\ % TEST with  -6.1
%&2 &  $-12.5 \pm 8.9$ & 190/192 & - & - & - &                                              953402.688 \\ % TEST with +17.0
%&2 &  $-12.7 \pm 9.2$ & 190/192 & - & - & - &                                              953275.562 \\ % TEST with  20.1 mag

\\
& \multicolumn{6}{c}{ $17 < G < 22.5$, $G$ and $(G_\text{BP} - G_\text{RP})$ corrected according to Eqs.~\ref{Eq-cor}-\ref{Eq-cor1}   } \\ %FINAL -6.0 +16.0 Gmag;  +3.5 Bp-Rp
35 & 0 &  $-12.4 \pm 3.4$ &  12/12   & - & - & - \\ % &                                            951245.750  \\ % nsim=5001   
36 & 1 &  $-13.4 \pm 7.3$ &  48/48  & - & - & - \\ % &                                             951503.938  \\ % nsim=1001  
37 & 2 &  $-13.0 \pm 9.1$ & 190/192 & - & - & - \\ % &                                             953260.438  \\ % nsim=1001  .
38 & 3 &  $-13.1 \pm 12.2$ & 705/768 & - & - & -\\ %  &                                            960330.062  \\ % nsim= 101 .
39 & 4 &  $-13.5 \pm 19.7$ & 2599/3072 & - & - & - \\ % &                                          989613.062  \\ % nsim=  15  4766m44.

\hline

\end{tabular} 
\tablefoot{
The result of the fitting Eq.~\ref{Eq-ZP2} to the QSO sample.
The value for $\langle C_0 \rangle$ is the median and dispersion over all HEALPix pixels with 40 objects per pixel or more based on
Monte Carlo simulations.
The number of these pixels is listed in column~4.
}
\label{Tab:Res}
\end{table*}

\begin{table} 
\caption{Example of PZPO and error over the HEALPix pixels.}

\begin{tabular}{rrrr} \hline \hline 
 HEALPix pixel & value  & error & Number \\
 \\
 \hline
 0 & -6.55 & 9.00 & 1308 \\ 
 1 & -20.79 & 10.05 & 488 \\ 
 2 & -18.46 & 6.86 & 838 \\ 
 3 & -16.25 & 12.93 & 635 \\ 
 4 & -28.44 & 15.57 & 592 \\ 
 5 & -8.94 & 16.20 & 453 \\ 
 6 & -21.67 & 12.68 & 647 \\ 
 7 & -22.48 & 13.13 & 515 \\ 
 8 & -27.29 & 8.09 & 956 \\ 
 9 & -34.50 & 14.58 & 656 \\ 
 10 & -40.48 & 9.50 & 2146 \\ 
   \ldots & \ldots & \ldots & \ldots \\
 760 & -8.36 & 9.88 & 550 \\ 
 761 &  3.38 & 8.43 & 562 \\ 
 762 & -8.08 & 12.00 & 404 \\ 
 763 &  0.46 & 10.69 & 542 \\ 
 764 &  5.99 & 11.31 & 486 \\ 
 765 & -9.45 & 8.54 & 776 \\ 
 766 & 11.18 & 9.30 & 591 \\ 
 767 & -8.84 & 8.87 & 775 \\ 
%768 &      5.44 &      0.53 &    - \\ 
%769 &      0.35 &      0.12 &    - \\ 
%770 &     -3.08 &      1.37 &    - \\ 
\hline

\end{tabular} 
\tablefoot{
Example of PZPO and error over the HEALPix pixels for level 3.
The files (corresponding to models 30-39 from Table~\ref{Tab:Res}) are available through the CDS.
The file contains the HEALPix pixel number, the PZPO with error (in \muas),
and column~4 is the number of objects in that pixel.
%The last entries (without value in the last column) are the parameters $C_1$, $C_2$, and $C_3$, when calculated.
}
\label{Tab:H768p3}
\end{table}

\begin{table} 
\caption{Properties of the solutions.}

\begin{tabular}{crrlll} \hline \hline 
  HEALPix & pixels & number of ill  & range in error \\% & S/N \\% & pixels with a small spatial PZPO \\
  level   &        & defined pixels &    (\muas)     \\% &    \\%  &        \\
 \hline
  0       &   12   &   0            &  0.63 - 1.74   \\% & 1.8 - 25   \\% & none \\
  1       &   48   &   0            &  1.16 - 10.8   \\% & 0.15 - 16   \\%&  3: $-0.73 \pm 3.3$,  12: $-0.39 \pm 2.6$,  47: $-0.64 \pm 2.5$ \\
  2       &  192   &   0            &  1.99 - 41     \\% & 0.04 - 8.9  \\%& 74: $+0.54 \pm 4.1$, 125: $+0.82 \pm 7.8$, 170: $+0.19 \pm 5.2$ \\

  3       &  768   &  12            &  3.23 - 82     \\% & 0.007 - 7.7 \\% &  63: $-0.79 \pm 6.5$, 140: $-0.37 \pm 6.8$, 192: $-0.97 \pm  7.7$, \\
%          &        &                &                 &             \\% & 445: $-0.97 \pm 5.7$, 208: $+0.83 \pm 6.0$, 248: $-0.91 \pm  7.1$,  \\
%          &        &                &                 &             & 286: $-0.25 \pm 9.4$, 294: $+0.69 \pm 9.9$, 314: $+0.08 \pm 8.4$, \\
%          &        &                &                 &             & 373: $-0.26 \pm 9.0$, 411: $-0.42 \pm 6.3$, 392: $-0.30 \pm 8.1$, \\
%          &        &                &                 &             & 439: $-0.06 \pm 7.6$, 444: $-0.78 \pm 7.5$, 599: $-0.60 \pm 7.1$, \\
%          &        &                &                 &             & 713: $+0.58 \pm  6.7$   \\

  4       & 3072   &  157          & 1.93 - 72   \\%  & 0.002 - 18  \\% &  366: $+0.86 \pm 5.1$, 372: $-0.12 \pm 9.3$, 485: $+0.50 \pm  9.5$, \\
%          &        &               &               &             &  498: $-0.38 \pm 9.5$, 589: $+0.48 \pm 9.8$, 627: $-0.70 \pm  9.5$, \\
%          &        &               &               &             &  723: $-0.40 \pm 6.3$, 937: $-0.31 \pm 5.9$, 1093: $-0.86 \pm 9.6$, \\
%          &        &               &               &             & 1255: $-0.34 \pm 8.9$, 1291: $-0.27 \pm 7.9$, 1481: $+0.90 \pm 4.1$, \\
%          &        &               &               &             & 1716: $+0.80 \pm 9.7$, 2031: $+0.10 \pm 8.2$, 2165: $-0.44 \pm 6.1$, \\
%          &        &               &               &             & 2399: $-0.12 \pm 8.1$, 2904: $-0.10 \pm 9.8$ \\

\hline

\end{tabular} 
\tablefoot{
 Column~1: HEALPix level,
  Column~2: total number of pixels on the Sky, 
  Column~3: number of pixels with 0 or 1 object,
  Column~4: range of the error in the PZPO for the pixels with 40 or more objects,
%  Column~5: S/N as $\mid$PZPO$\mid$ divided by the error therein for the pixels with 40 or more objects.
%  Column~6: HEALPix pixel number, PZPO and error (in \muas), for the pixels with $\mid$PZPO$\mid< 1$~\muas\ and error $< 10$~\muas.
%
}
\label{Tab:Sum}
\end{table}

\subsection{Wide binaries}
\label{SS-WBR}

\begin{figure}
\centering

\begin{minipage}{0.47\textwidth}

\resizebox{\hsize}{!}{\includegraphics{ParGmag_all_593401.ps}}

\resizebox{\hsize}{!}{\includegraphics{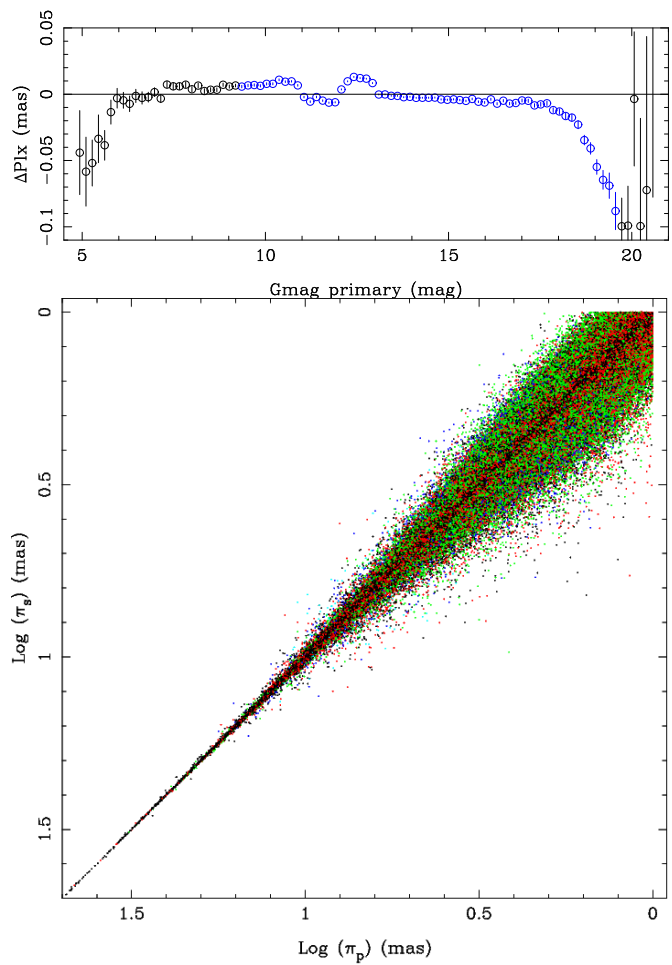}}

\end{minipage}

\caption{Top panel. Parallax difference between primary and secondary component in wide binaries, as a function of
  primary $G$ magnitude. Bins with more than 1000 objects are plotted in blue.
Bottom panel. Parallax of the secondary binary component plotted against that of the primary.
Objects with primary $G$ magnitude larger than 5, 15, 17, 19, and 20~mag are plotted as
black, red, green, dark blue, and light blue dots, respectively.
Plotted are about 590~000 binaries where both components pass the criteria on GOF and RUWE.
}
\label{Fig:WB1}
\end{figure}

The top panel in Fig.~\ref{Fig:WB1} shows a binned version of the parallax difference between the primary and secondary component
as a function of primary $G$ in the top panel.
A similar diagrams was shown in \cite{Fabricius20} (their Fig.~22) %% and \citet{GEDR3_LindegrenZP} 
What is striking  is the sharp decrease of the parallax difference for faint magnitudes.
This is due to selection effects in the \cite{ElBadry21} sample. Readily visible in the bottom panel are the conditions
$\pi_{\rm p} >1$ and $\pi_{\rm s} >1$~mas that were imposed (but there are others on the (relative) parallax accuracy,
see their section~2), and that for a given $\pi_{\rm p}$ there are
many more objects with $\pi_{\rm s} < \pi_{\rm p}$ than the inverse.

\begin{figure}
\centering

\begin{minipage}{0.47\textwidth}

\resizebox{\hsize}{!}{\includegraphics{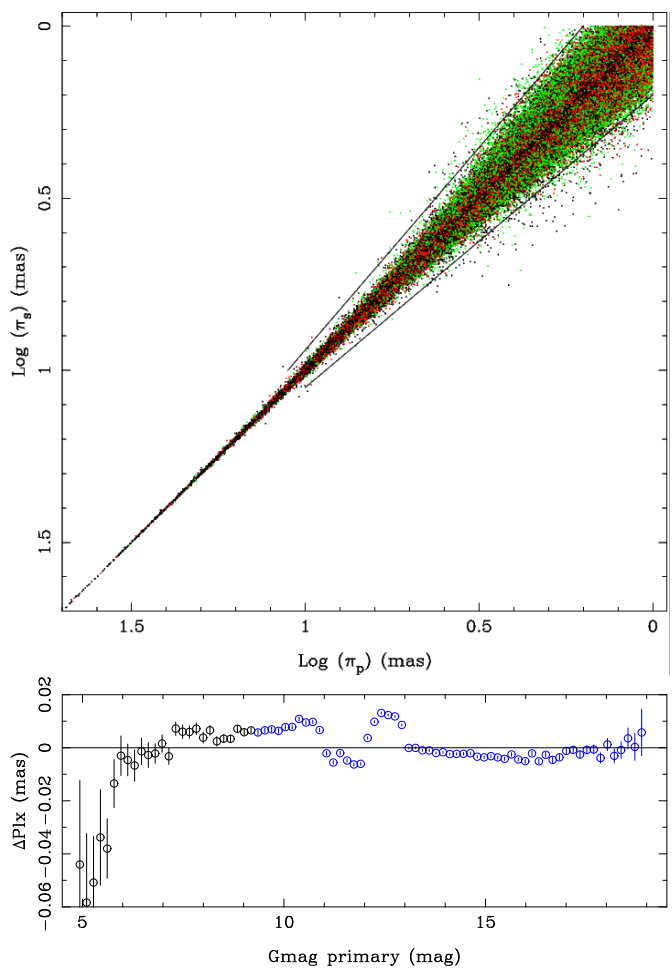}}
  
\resizebox{\hsize}{!}{\includegraphics{ParGmag_final_479448.ps}}

\end{minipage}

\caption{As Fig.~\ref{Fig:WB1} for $G_{\rm s}$ restricted to  $< 19.0$~mag. 
  After eliminating $\sim$700 extreme outliers (those outside the two plotted lines in the top panel),
  a sample of about 480~000 objects remain. The bottom panel shows the parallax difference between primary and secondary
  component for that sample as a function of $G_{\rm p}$. Note the different range in the ordinate compared to Fig.~\ref{Fig:WB1}.
}
\label{Fig:WB2}
\end{figure}

Restricting the magnitudes to less than 19~mag seems to largely remove this asymmetric behaviour (top panel Fig.~\ref{Fig:WB2}) and
also removes the unexpected tendency of the parallax difference as a function of magnitude (bottom panel).

\begin{figure}
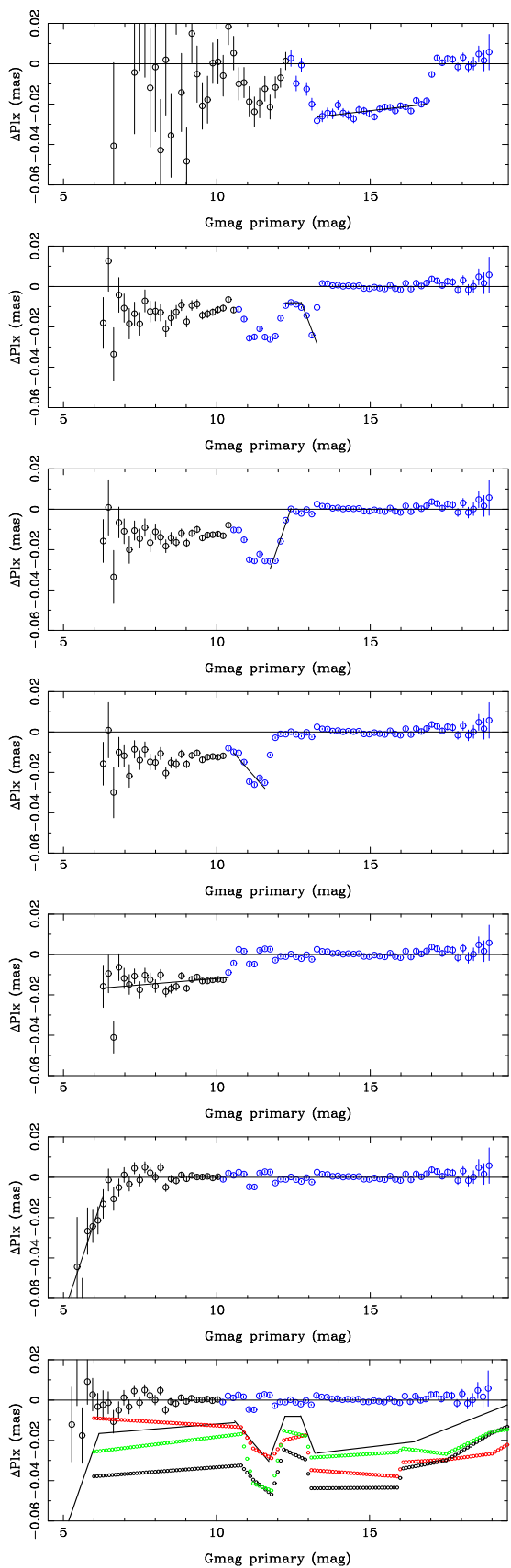

\centering

\begin{minipage}{0.39\textwidth}

\resizebox{\hsize}{!}{\includegraphics{ParGmag_GmagCorrected_Gslimited13to22.ps}}

\resizebox{\hsize}{!}{\includegraphics{ParGmag_GmagCorrected_Gslimited12p2to13p2.ps}}

\resizebox{\hsize}{!}{\includegraphics{ParGmag_GmagCorrected_Gslimited11p7to12p4.ps}}

\resizebox{\hsize}{!}{\includegraphics{ParGmag_GmagCorrected_Gslimited10p5to11p6.ps}}

\resizebox{\hsize}{!}{\includegraphics{ParGmag_GmagCorrected_Gslimited6p3to10p4.ps}}

\resizebox{\hsize}{!}{\includegraphics{ParGmag_GmagCorrected_Gslimited5p3to6p3.ps}}

\resizebox{\hsize}{!}{\includegraphics{ParGmag_GmagCorrected_Gslimited5p0to22.ps}}

\end{minipage}

\caption{As Fig.~\ref{Fig:WB1} with parallaxes corrected according to Eq.~\ref{Eq-corWB} in consecutive steps (see main text).
  The bottom panel shows the applied correction (Eq.~\ref{Eq-corWB}) as the black line.
  The small black, red, and green circles represent the L20 correction for $\beta= 0, -60,$ and $+60\degr$.
  There is an offset as the L20 corrections are absolute, while the corrections applied to the WB sample are relative
  to the correction at $G$ = 20~mag.
}
\label{Fig:WB3}
\end{figure}

It is now possible to iteratively study the PZPO based on WBs as a function of $G$ magnitude.
The first step is to correct the parallaxes according to Eq.~\ref{Eq-cor}. As the $G$ magnitude of the binary sample is limited to
$G$ = 19~mag this implies correcting the parallaxes of all objects with $G >$17 by +6.0 \muas/mag.
One can then plot the parallax difference against magnitude, only considering secondaries fainter than 17~mag.
The top panel of Fig.~\ref{Fig:WB3} shows the result. The PZPO is essentially independent of $G$ at the faint end, the
weighted mean of the 11 bins fainter than 17.1~mag is $1.6 \pm 0.6$~\muas. In the range between
$\sim$13.3 and $\sim$17~mag the PZPO can be well approximated by a linear behaviour as indicated by the black line.
In a second step this offset can be applied as well in this magnitude range, and the PZPO can be determined using
secondaries fainter than $13.3$~mag.
The consecutive panels in Fig.~\ref{Fig:WB3} show how this procedure can be applied to brighter and brighter magnitudes.
The bottom panel shows the final result.
The weighted mean of the residuals is 0.05~\muas\ with an rms of 2.7~\muas\ (for $G> 6$), 4.3~\muas\ (for $G> 5$), and
13~\muas\ (for $5< G < 6$~mag). % including colour correction -0.82 2.61 4.27 12.7
The corrections that were applied with the range of $G$ magnitudes determined so that the correction is continuous in
$G$ is given by Eq.~\ref{Eq-corWB}. It is show as the black line in the bottom panel.

The bottom panel of Fig.~\ref{Fig:WB3} also shows for comparison the correction by L20 for $\nu_{\rm eff} = 1.55$
(corresponding to $(G_\text{BP} - G_\text{RP})$ = 0.8~mag) and $\beta = 0\degr$ (small black circles), $+60\degr$ (green), and $-60\degr$ (red circles).
The behaviour for $\beta = 0\degr$ of the L20 correction is very similar in shape and amplitude to the correction derived here.
There is an offset due to the fact that the L20 corrections are absolute while the corrections in Eq.~\ref{Eq-corWB} are
relative to the correction at $G$ = 20 which is $\approx -12.6$~\muas\ (see Table~\ref{Tab:Res}) which is indeed 
about the difference at the faint end.
What is remarkable is that the L20 correction also depends in a particular way on ecliptic latitude.
%1
%
For bright magnitudes the L20 correction for $-60\degr$ lies above that for $+60\degr$, while for $G$ \more 13~mag it is the
inverse. In addition, the dependence on $\beta$ increases with brighter magnitudes.

\begin{equation}
\Delta\pi=\begin{cases}
$\phantom{1}$+6.0  \; (G - 19.900)   $\phantom{00}$+0.000 & 16.450 \leq G < 19.900 \\
+1.78              \; (G - 13.265)                -26.372 & 13.218 \leq G < 16.450 \\
-40.2              \; (G - 12.755)   $\phantom{00}$-7.823 & 12.761 \leq G < 13.218 \\
$\phantom{-99.9    \; (G - 19.999)}$ $\phantom{00}$-8.064 & 12.243 \leq G < 12.761  \\ 
+42.3              \; (G - 11.735)                -29.531 & 11.713 \leq G < 12.243    \\ 
-17.2              \; (G - 10.545)                -10.366 & 10.591 \leq G < 11.713  \\ 
+1.26              \; (G - $\phantom{00}$6.295)   -16.579 & $\phantom{0}$6.162 \leq G < 10.591   \\ 
+57.3              \; (G - $\phantom{00}$5.275)   -67.589 & $\phantom{0}$5.275 \leq G < $\phantom{1}$6.162   \\
\end{cases}
  \label{Eq-corWB}
\end{equation}

As error in this parallax correction a 1~\muas\ systematic error is added in quadrature to a random error
of 2.7~\muas\ for $G > 6$ and 13~\muas\  for  $G \le 6$.

\section{ Application of the PZPO correction}
\label{S-Appl}

In this section the results are being discussed related to the application of the PZPO correction.

\subsection{QSOs }
\label{SS-QSO}

The first application of the PZPO correction is to the QSO sample itself and a comparison to using the L20 correction.
Table~\ref{Table:QSOCorr} provides the L20 correction, and the spatial, magnitude and total correction in the present work, as well as the
offset after applying the L20 correction and the correction in the present work for the different HEALPix levels.
The weighted mean and the error therein are quoted in all cases. As the L20 correction comes without an error one has been assigned.
It has been chosen to be a constant such that the error in the weighted mean of the L20 and the correction in
the present work (Cols~2 and 5) are the same for HEALPix level 0 and equals 3.0~\muas. This choice has no practical impact.
The error in the weighted mean after applying the corrections (Cols~6 and 7) is independent of this choice, and is actually virtually the same
for both type of corrections ($\sim$0.3~\muas), as this error is dominated by the error in the observed parallaxes.

\begin{figure}
\centering

\begin{minipage}{0.47\textwidth}
\resizebox{\hsize}{!}{\includegraphics{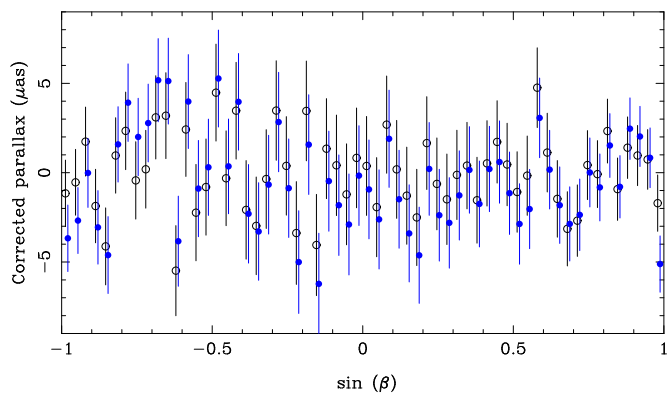}}
\end{minipage}

\caption{Residual in the observed parallax after applying the correction in the present work (in black open circles, offset by $-0.004$ units horizontally)
  and that in L20 (in blue filled circles, offset by $+0.004$ units) versus $\sin \beta$ for the QSO sample. Sixty bins have been used,
  and HEALPix level 2 has been used in the calculations.
}
\label{Fig:QSO_ELAT}
\end{figure}

The results in Table~\ref{Table:QSOCorr} give the overall comparison for $\sim$821~000 QSOs in the sample, but as the main difference between the approach
in L20 and in the present work is in the dependence of the correction on sky position this dependence is of interest.
Figure~\ref{Fig:QSO_ELAT} shows the corrected parallax after applying the correction in the present work (in black) and
that in L20 (in blue) using 60 bins in $\sin \beta$. The black points tend to be closer to the line of zero offset and Table~\ref{Table:QSOCorrELAT}
contains the details for the different HEALPix levels.
The table lists the median over the bins and the scatter around the median (calculated as 1.4826$\cdot$MAD), and shows that the scatter decreases
with increasing spatial resolution when using the present correction. 
As shown in the next two subsections, this, will not be the case in general however.
As the sample to define the spatial correction is the same as to which it is applied there are no undefined spatial pixels being used.
In general, increasing the spatial resolution (increasing the HEALPix level) will lead to an increasing number of stars to be in spatial pixels
that are undefined (insufficient number of QSOs), so that there will be an optimal HEALPix level to be used.

\begin{table*} 
\setlength{\tabcolsep}{1.6mm}
\caption{Parallax corrections for the QSO sample}

\begin{tabular}{ccccccccl} \hline \hline

  HEALPix  &  L20         & PW spatial    & PW magnitude     & PW total          & $\Delta$(PZPO corrected   & $\Delta$(PZPO corrected \\
  level    &  correction  & correction    & correction       & correction        &  parallax, L20)            &   parallax, PW)   \\
           &  (\muas)     &  (\muas)      & (\muas)          &  (\muas)          &   (\muas)                 &   (\muas)    \\ \hline
%           & \multicolumn{7}{c}{ 66 CCs following \citet{RiessGEDR3}} \\
 0 & $-17.11 \pm 0.0034$ & $-13.49 \pm 0.0010$ & $-5.04 \pm 0.0032$ & $-18.66 \pm 0.0034$ & $-0.502 \pm 0.295$ & $+0.008 \pm 0.295$ \\
 1 & $-17.11 \pm 0.0034$ & $-13.25 \pm 0.0019$ & $-5.04 \pm 0.0032$ & $-18.73 \pm 0.0038$ & $-0.502 \pm 0.295$ & $+0.016 \pm 0.295$ \\
 2 & $-17.11 \pm 0.0034$ & $-13.31 \pm 0.0036$ & $-5.04 \pm 0.0032$ & $-18.70 \pm 0.0050$ & $-0.502 \pm 0.295$ & $+0.002 \pm 0.295$ \\
 3 & $-17.11 \pm 0.0034$ & $-13.38 \pm 0.0070$ & $-5.04 \pm 0.0032$ & $-18.68 \pm 0.0079$ & $-0.502 \pm 0.295$ & $-0.007 \pm 0.295$ \\
 4 & $-17.11 \pm 0.0034$ & $-14.08 \pm 0.0104$ & $-5.04 \pm 0.0032$ & $-19.16 \pm 0.0114$ & $-0.502 \pm 0.295$ & $+0.005 \pm 0.296$ \\
 \hline
 \end{tabular} 
\tablefoot{
Column~1 gives the HEALPix level,
column~2 gives the weighted mean and error of the L20 correction,
columns~3-5 give the weighted mean and error for the spatial correction, the magnitude correction, and the total correction of the present work (PW),
Column~6-7 give the weighted mean and error after applying the correction in L20 and of the present work, respectively.
}
\label{Table:QSOCorr}
\end{table*}

\begin{table} 
\caption{Parallax corrections for the QSO sample when binned against ecliptic latitude}

\begin{tabular}{ccccccccl} \hline \hline

  HEALPix  &   $\Delta$(PZPO corrected   & $\Delta$(PZPO corrected \\
  level    &   parallax, L20)            &   parallax, present work)   \\
           &     (\muas)                 &   (\muas)    \\ \hline
 0 &  $-0.82 \pm 2.75$ & $-0.086 \pm 3.06$ \\
 1 &  $-0.82 \pm 2.75$ & $+0.078 \pm 2.50$ \\
 2 &  $-0.82 \pm 2.75$ & $-0.080 \pm 2.09$ \\
 3 &  $-0.82 \pm 2.75$ & $-0.166 \pm 1.80$ \\
 4 &  $-0.82 \pm 2.75$ & $-0.090 \pm 1.59$ \\
 \hline
 \end{tabular} 
\tablefoot{
Column~1 gives the HEALPix level,
column~2-3 give the median and 1.4826$\cdot$MAD using the L20 correction and the present work, respectively.
}
\label{Table:QSOCorrELAT}
\end{table}

\subsection{Independent trigonometric parallaxes}
\label{SS-Ind}

\begin{figure}
\centering

\begin{minipage}{0.47\textwidth}
\resizebox{\hsize}{!}{\includegraphics{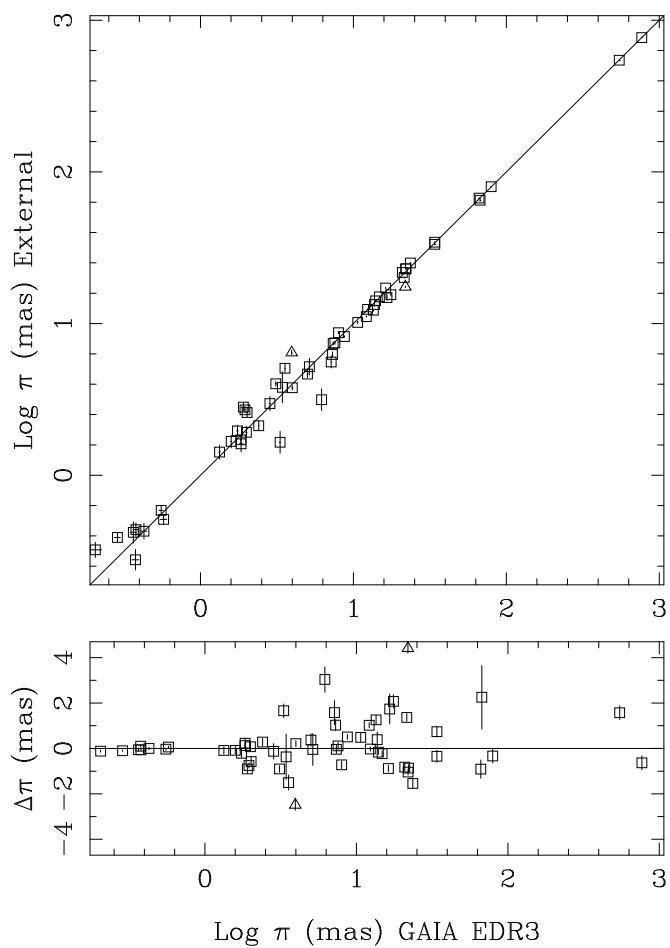}}
\end{minipage}

\caption{Independent trigonometric parallax plotted against GEDR3 parallax.
  The bottom panel displays the residual, where the error bar in the ordinate combines the error in the {\it Gaia} and
  the independent parallax in quadrature.
  Two stars where the residual is more than ten times the combined error bar are plotted as open triangles.
}
\label{Fig:FGS_ALL}
\end{figure}

In Sect.~\ref{SS-FGS} a sample of 111 stars with independent trigonometric paralax data was introduced (Table~\ref{Tab-Targets}) of
which 57 pass the selection on GOF and RUWE. Figure~\ref{Fig:FGS_ALL} compares these parallaxes to the GEDR3 ones in
the top panel, while the residuals are shown in the bottom panel.

Two stars are excluded in the further analysis, VY Pyx and HD~285876.
\citet{Benedict17} mention that VY Pyx is an outlier, lying 1.19~mag of the PL-relation they derived.
Adopting the \G\ parallax would shift this object 1.06~mag closer, and hence in agreement, with the PL-relation.
Although \citet{Benedict17} carefully analysed all steps in their procedure, it is likely that the FGS parallax is in error.
For vA~645 (HD~285876) the difference between \G\ and FGS parallax is 20$\sigma$, much larger than one can reasonably
attribute to a statistical outlier.

\begin{figure}
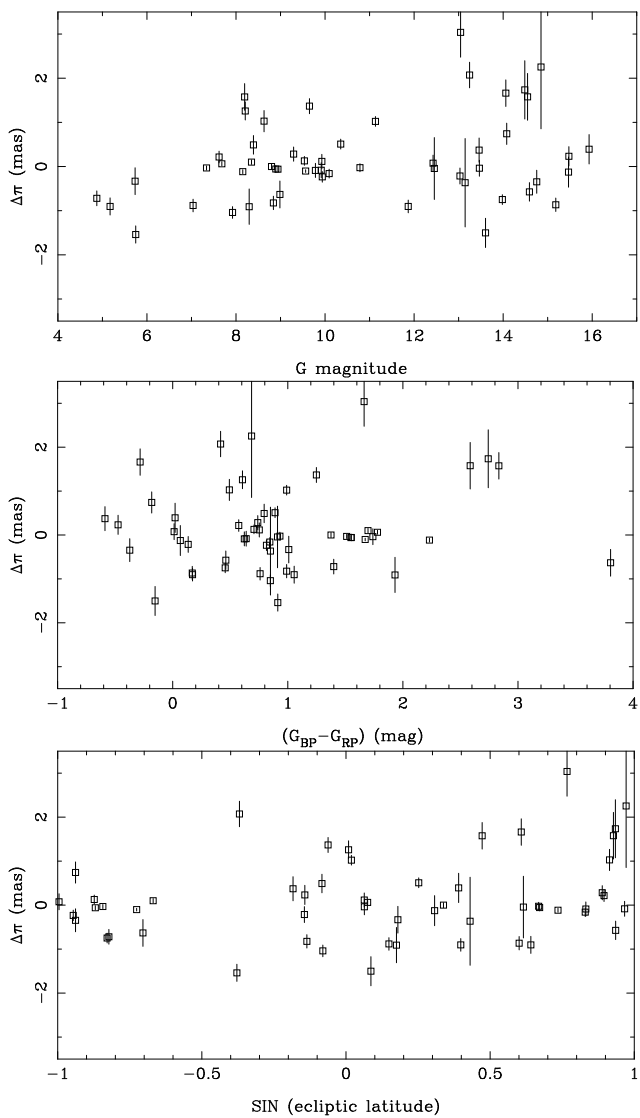

\centering

\begin{minipage}{0.45\textwidth}
\resizebox{\hsize}{!}{\includegraphics{ParOff_Gmag_FGS_minus2.ps}}
\end{minipage}
\begin{minipage}{0.45\textwidth}
\resizebox{\hsize}{!}{\includegraphics{ParOff_BpRp_FGS_minus2.ps}}
\end{minipage}
\begin{minipage}{0.45\textwidth}
\resizebox{\hsize}{!}{\includegraphics{ParOff_Elat_FGS_minus2.ps}}
\end{minipage}

\caption{Difference between the independent trigonometric parallax and the GEDR3 parallax plotted against $G$,
  $(G_\text{BP} - G_\text{RP})$ colour, and $\sin \beta$.
  The two outliers mentioned in Fig.~\ref{Fig:FGS_ALL} have been removed.
}
\label{Fig:FGS_minus2}
\end{figure}

Figure~\ref{Fig:FGS_minus2} plots the residuals against $G$ magnitude, $(G_\text{BP} - G_\text{RP})$ colour and $\sin \beta$.
Although these are the objects with the best independent trigonometric parallaxes the error bars on the differential parallax
are dominated by the error in the external parallax and the range in the ordinate ($\sim$4~mas) is
much larger than when intercomparing GEDR3 parallaxes where differences in parallax due to sky position,
magnitude and colour are of order 100 times smaller ($\sim$0.04~mas, e.g. Figs.~\ref{Fig:QSOparoff} or \ref{Fig:WB3}).
This is probably the reason that no trends are obvious.

Table~\ref{Table:FGS} gives the median and weighted mean with error of the difference between observed
and the independent trigonometric parallax.
The first five entries are for the entire sample applying increasingly stricter selection criteria.
The last two entries are specifically for the CCs, Type-{\sc ii} cepheids (T2C) and RRL stars in the sample.
These objects are of special interest to the distance scale, and they are all radially pulsating stars of
similar magnitude (7~\less~$G$~\less~10) and colour (0.5~\less~$(G_\text{BP} - G_\text{RP})$~\less~ 2.3~mag).
Appendix~\ref{AppCC} gives some more details on this subsample.

Table~\ref{Table:FGSCorr} provides the spatial, magnitude and total correction, as well as the
offset between the corrected GEDR3 and  independent trigonometric parallaxes.
This is done for three representative samples, and for the different HEALPix levels.
Ideally, the weighted mean of the difference between the corrected GEDR3 parallax and the  independent trigonometric parallax (Col.~5) should
be zero within the error bars, and this is indeed the case. However, some trends are observed. For the larger HEALPix levels an increasing number
of stars will be missing and this results in a marked increase in the scatter. On the other hand one should favour the best mapping of the
spatial variations. For the samples discussed here this would imply using the results for HEALPix level 2 as the most appropriate.
However this choice will depend on the properties of the external sample (number of stars, distribution on the sky, accuracy of the external parallaxes).

\begin{table*} 
\setlength{\tabcolsep}{1.5mm}
\caption{PZPOs for the sample with  independent trigonometric parallaxes}

\begin{tabular}{cccl} \hline \hline
  
  median   & weighted mean and error  & N &  description        \\
  (\muas)  & (\muas)                  &   &                     \\ \hline
 $-39 $ & $-35 \pm 14$ & 57 & all that pass the GOF/RUWE selection       \\ % \pm 786
 $-39 $ & $-45 \pm 14$ & 55 & excluding VY Pyx and vA 645 (10$\sigma$ outliers) \\ % \pm 637
 $-45 $ & $-71 \pm 14$ & 53 & excluding 8$\sigma$ outliers \\ % \pm 618
 $-39 $ & $-36 \pm 14$ & 46 & excluding 6$\sigma$ outliers \\ % \pm 434
 $-39 $ & $-39 \pm 14$ & 39 & excluding 6$\sigma$ outliers, $\sigma_{\pi} <0.4$~mas \\ % \pm 377
  \\
 $-33 $ & $-31 \pm 16$ & 15 & all CCs, T2C, RRL that pass the GOF/RUWE selection (and excluding VY Pyx as well) \\ % \pm 124
 $-33 $ & $-29 \pm 16$ & 13 & as above, excluding Polaris B and FF Aql as well \\ % \pm 104
 \hline
 \end{tabular} 
\tablefoot{  
Columns~1 and 2 list the median and weighted mean offset between the GEDR3 parallax and the independent trigonometric parallax
  for the sample discussed in Sects.~\ref{SS-FGS} and \ref{SS-Ind}.
Column~3 lists the number of stars, and 
column~4 gives a description of the selection criteria.
}
\label{Table:FGS}
\end{table*}

\begin{table*} 
\caption{ Parallax corrections for the sample with independent trigonometric parallaxes}

\begin{tabular}{cccccc} \hline \hline
  
  HEALPix  & spatial correction & magnitude correction & total correction & $\Delta$ (PZPO corrected parallax) & N      \\
  level    & (\muas)            & (\muas)              &  (\muas)         & (\muas)            &        \\ \hline
           & \multicolumn{4}{c}{Sample: 55 stars that pass GOF/RUWE selection} \\
  0 & $-13.18 \pm 0.14$ & $-17.47 \pm  0.40$ &  $-30.10 \pm 0.44$ & $-11.17 \pm  13.88$ & 54 \\
  1 & $-14.67 \pm 0.29$ & $-17.47 \pm  0.40$ &  $-32.11 \pm 0.53$ & $-13.03 \pm  13.93$ & 54 \\
  2 & $-13.60 \pm 0.58$ & $-17.47 \pm  0.40$ &  $-32.07 \pm 0.78$ & $\phantom{ }-7.69 \pm  14.16$ & 54 \\
  3 & $-10.72 \pm 1.12$ & $-17.58 \pm  0.41$ &  $-30.35 \pm 1.26$ & $\phantom{ }+2.46 \pm  18.63$ & 53 \\
  4 & $-16.11 \pm 1.46$ & $-18.07 \pm  0.44$ &  $-35.74 \pm 1.65$ & $\phantom{ }+2.40 \pm  23.03$ & 45 \\
  \\
    & \multicolumn{4}{c}{Sample: 39 stars that pass GOF/RUWE selection, excluding 6$\sigma$ outliers, $\sigma_{\pi} <0.4$~mas } \\
  0 & $-13.67 \pm 0.19$ & $-16.68 \pm  0.48$ &  $-29.50 \pm 0.53$ & $\phantom{ }-4.86 \pm  14.50$ & 38 \\
  1 & $-14.94 \pm 0.37$ & $-16.68 \pm  0.48$ &  $-31.55 \pm 0.65$ & $\phantom{ }-6.85 \pm  14.55$ & 38 \\
  2 & $-13.98 \pm 0.75$ & $-16.68 \pm  0.48$ &  $-31.95 \pm 0.99$ & $\phantom{ }-2.16 \pm  14.81$ & 38 \\
  3 & $-12.58 \pm 1.53$ & $-16.81 \pm  0.49$ &  $-31.56 \pm 1.70$ & $+13.81 \pm 20.16$ & 37 \\
  4 & $-25.05 \pm 1.94$ & $-17.58 \pm  0.54$ &  $-43.10 \pm 2.24$ & $\phantom{ }-4.42 \pm  25.71$ & 30 \\
\\
           & \multicolumn{4}{c}{Sample: 13 CCs, T2C, RRL} \\
  0 & $-13.91 \pm 0.35$ & $-13.47 \pm  0.80$ &  $-26.72 \pm 0.88$ & $\phantom{ }-6.75 \pm  15.91$ & 13 \\
  1 & $-14.49 \pm 0.70$ & $-13.47 \pm  0.80$ &  $-27.49 \pm 1.12$ & $\phantom{ }-2.87 \pm  15.97$ & 13 \\
  2 & $-14.31 \pm 1.41$ & $-13.47 \pm  0.80$ &  $-28.13 \pm 1.85$ & $\phantom{ }+3.92 \pm  16.30$ & 13 \\
  3 & $-10.48 \pm 3.19$ & $-13.47 \pm  0.80$ &  $-24.47 \pm 3.56$ & $\phantom{ }+2.70 \pm  23.70$ & 13 \\
  4 & $ -7.38 \pm 5.54$ & $-12.95 \pm  1.01$ &  $-20.13 \pm 5.79$ & $+16.91 \pm  32.86$           &  \phantom{1}8 \\
 \hline
 \end{tabular} 
\tablefoot{
Column~1 gives the HEALPix level considered (defining the  spatial correction term),
columns~2-5 give the weighted mean and error for the spatial correction, the magnitude
correction, the total correction, and the  offset between the corrected GEDR3 parallax and the external parallax.
Column~6 gives the number of stars. The samples refer to those defined in Tab.~\ref{Table:FGS}.
}
\label{Table:FGSCorr}
\end{table*}

\subsection{Classical cepheids}
\label{SS:CC}

As a second application of the spatial and magnitude corrections derived in the present paper the sample of Galactic CCs of \citet{RiessGEDR3} is studied.
They discuss a sample of 75 CCs with HST photometry which is used to calibrate the extragalactic distance scale along the lines outlined in earlier
works \citep{Riess16, RiessGDR2, Riess19}. They correct the GEDR3 parallaxes using the L20 formalism and fit the slope, zero point and
metallicity dependence of the PL relation as well as a constant offset between the photometric parallaxes and the corrected GEDR3 values.
Fits where some of these parameters are fixed are also presented.
In their analysis they increased the GEDR3 parallax uncertainties by 10\%, which is not done here.
One important conclusion in the present context is that \citet{RiessGEDR3} find that the L20 procedure over corrects the PZPO by $14 \pm 6$~\muas. 

Table~\ref{Table:CC} contains the result of the calculations for two samples.
The first is the sample of 66 stars retained by \citet{RiessGEDR3}. This is the full sample of 75 stars minus 9 stars excluded in their best fit analysis.
Six were excluded there because their GOF $>$ 12.5 (SV Per, RW Cam, U Aql, DL Cas, SY Nor, RX Cam), one, CY Aur, because it is an outlier
in the L20 correction, and S Vul and SV Vul that are marginal outliers of the PL-relation\footnote{It was confirmed (Riess 2021, private communication)
that two corrections are necessary in Table 1 of \citet{RiessGEDR3} to match their analysis; Z Sct is missing there but is available in
Table~1 of \citealt{RiessGDR2}, and the high GOF flag on AD Pup (GOF = 12.48) should instead appear on RX Cam (GOF = 28.7).}.
The second sample are the 54 stars that remain after the applying the criteria on GOF and RUWE used in this paper.

The reference parallax is the photometric parallax (with error) from Table~1 in \citet{RiessGEDR3} that is derived from
the HST photometry, the pulsation period, and the PL relation from \citet{Riess16, Riess19}.

\begin{figure}
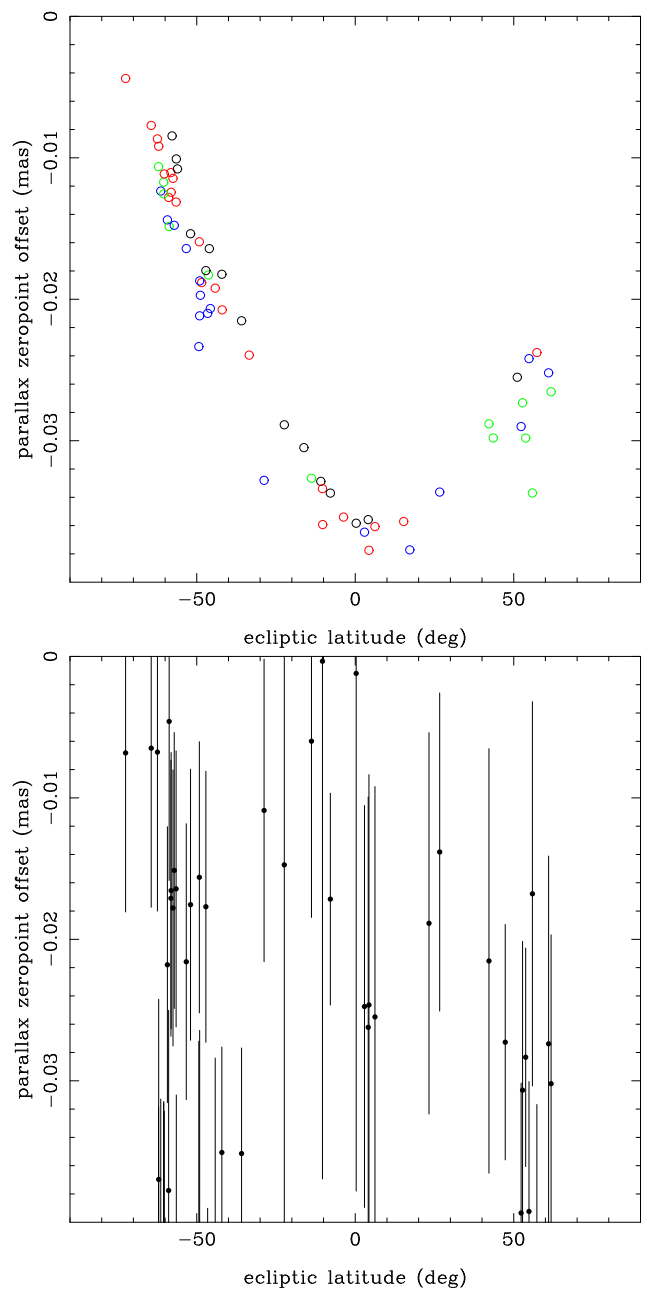

\centering

\begin{minipage}{0.45\textwidth}
\resizebox{\hsize}{!}{\includegraphics{Correction_Cep_66_L20.ps}}
\end{minipage}
\begin{minipage}{0.45\textwidth}
\resizebox{\hsize}{!}{\includegraphics{Correction_Cep_54_HLP2.ps}}
\end{minipage}

\caption{Top panel: PZPO correction by L20 for the sample of 66 CCs analysed by \citet{RiessGEDR3} (cf. their Fig.~2).
  The colours represent different ranges in $G$: black ($G \le 7$), red ($7 < G \le 8.5$), green ($8.5 < G \le 9.0$),
  and blue ($9 < G \le 11.5$).
  Bottom panel: Correction proposed here for the stricter selected sample of 54 stars at HEALPix level 2.
}
\label{Fig:CCs}
\end{figure}

Column~2 of Tab.~\ref{Table:CC} gives the weighted mean offset between the observed GEDR3 parallax and the photometric parallax.
It is unusually large ($-6.4$ to $-7.7$~\muas, see below).
Column~3 lists the weighted mean of the L20 correction, and the upper panel of Fig.~\ref{Fig:CCs} show the dependence on $\beta$.
A similar diagram was shown for the 75 CCs in \citet{RiessGEDR3}. What is striking is the close to parabolic shape of the correction
which is build-in in the L20 approach.
The other columns show the weighted mean spatial, magnitude, and total correction, the offset between the corrected GEDR3 parallax
and the photometric parallax, and the number of objects.
If the GEDR3 parallaxes are corrected by the L20 formalism (on a star-by-star basis) the weighted mean offset with the photometric parallax
becomes $+14.3 \pm 2.9$~\muas, consistent with \citet{RiessGEDR3}, and indicating an overcorrection by the L20 formalism.
By increasing the photometric parallaxes by $\sim$3.3\% one can obtain a weighted mean offset between the L20 corrected and photometric
parallax consistent with zero ($0.00 \pm 2.93$~\muas).
Such an increase is consistent with the result reported in the last entry of table~2 in \citet{RiessGEDR3} where they forced a fit without
additional PZPO to determine the zero point of the PL relation. The value reported there ($-5.865 \pm 0.013$) is consistent with finding here
that implies a zero point of $-5.93 + 5 \log 1.0325 = -5.861$.

The second block in Table~\ref{Table:CC} shows similar results for the smaller sample that fulfils the criteria on GOF and RUWE imposed here.
The bottom panel of Fig.~\ref{Fig:CCs} shows that there is no the dependence  of the total correction (with error bar) proposed
in the present paper on $\beta$.
It is remarked that the error in the average total correction (Col.~6) is similar or smaller than the average L20 correction (Col.~3)
  for HEALPix levels 0, 1, and 2.

Adopting the standard photometric parallax leads to overcorrection of $\sim$20~\muas.
Increasing the photometric parallax by a factor $1.0505 \pm 0.0080$ (implying a PL zero point of $-5.823 \pm 0.016$, and
$H_0 = 76.2 \pm 1.3$ km/s/Mpc) will
lead to weighted mean  offset between the corrected GEDR3 and the photometric parallax consistent with zero.
It also implies a weighted mean  offset of the observed GEDR3 and the photometric parallax of $-29 \pm 3$~\muas, which is
very similar to other bright ($G$~\less~10-11~mag) samples, for example the stars with external trigonometric parallaxes and the
subsample of pulsating stars ($-39 \pm 14$, respectively, $-29 \pm 16$~\muas\ from Tab.~\ref{Table:FGS}) or the sample of
EBs ($-37 \pm 20$~\muas, \citealt{StassunTorres21}) or WUMa-type EBs ($-28.6 \pm 0.6$~\muas\ for the 5-parameter solution, \citealt{Ren21}).

\begin{table*} 
\setlength{\tabcolsep}{1.2mm}
\caption{Parallax corrections for samples of Galactic CCs}

\begin{tabular}{ccccccccl} \hline \hline

  HEALPix  & $\Delta$(uncorrected & L20         & PW spatial     & PW magnitude  & PW total      & $\Delta$(PZPO corrected   & N & Remarks \\
  level    &  parallax)            & correction  & correction  & correction & correction &  parallax, PW)            &   &     \\
           & (\muas)              & (\muas)     &  (\muas)    & (\muas)    &  (\muas)   &   (\muas)    \\ \hline
           & \multicolumn{7}{c}{ 66 CCs following \citet{RiessGEDR3}} \\
  0 & $-6.36 \pm 2.83$ & $-22.1 \pm 1.23$ & $-14.10 \pm 0.17$ & $-14.55 \pm 0.35$ & $-27.82 \pm 0.39$ & $+22.0 \pm 2.87$ & 66 \\
  1 &                  &                  & $-14.13 \pm 0.38$ & $-14.55 \pm 0.35$ & $-27.57 \pm 0.55$ & $+20.3 \pm 2.90$ & 66 \\
  2 &                  &                  & $-14.43 \pm 1.25$ & $-14.55 \pm 0.35$ & $-28.94 \pm 1.31$ & $+25.1 \pm 3.26$ & 66 \\
  3 &                  &                  & $-10.43 \pm 4.12$ & $-14.57 \pm 0.36$ & $-25.34 \pm 4.17$ & $+17.4 \pm 6.73$ & 63 \\
  4 &                  &                  & $\phantom{ }-8.58 \pm 8.57$ & $-14.67 \pm 0.47$ & $-23.90 \pm 8.64$ & $+24.7 \pm 11.5$ & 38 \\
  \\
  2 & $-20.5 \pm 2.83$ &                  & $-14.43 \pm 1.25$ & $-14.55 \pm 0.35$ & $-28.94 \pm 1.31$ & $+10.4 \pm 3.26$ & 66 & $\pi_{\rm phot} \cdot 1.0325$\tablefootmark{a} \\
  \\
           & \multicolumn{7}{c}{ 54 CCs following the GOF/RUWE selection in the present work } \\
  0 & $-7.66 \pm 3.04$ & $-22.0 \pm 1.36$ & $-13.66 \pm 0.19$ & $-14.54 \pm 0.39$ & $-27.39 \pm 0.44$ & $+20.4 \pm 3.08$ & 54 \\
  1 &                  &                  & $-13.51 \pm 0.42$ & $-14.54 \pm 0.39$ & $-27.15 \pm 0.60$ & $+18.5 \pm 3.12$ & 54 \\
  2 &                  &                  & $-13.57 \pm 1.41$ & $-14.54 \pm 0.39$ & $-28.04 \pm 1.47$ & $+22.2 \pm 3.53$ & 54 \\
  3 &                  &                  & $\phantom{ }-8.05 \pm 4.66$ & $-14.57 \pm 0.40$ & $-22.70 \pm 4.73$ & $+13.4 \pm 7.62$ & 51 \\
  4 &                  &                  & $\phantom{ }-6.94 \pm 9.24$ & $-14.82 \pm 0.55$ & $-22.35 \pm 9.32$ & $+22.7 \pm 12.7$ & 27 \\
  \\
  2 & $-29.2 \pm 3.05$ &                  & $-13.57 \pm 1.41$ & $-14.54 \pm 0.39$ & $-29.22 \pm 3.05$ & $+0.00 \pm 3.53$ & 54 & $\pi_{\rm phot} \cdot 1.0505$ \\

 \hline
 \end{tabular} 
\tablefoot{
Column~1 gives the HEALPix level considered (defining the spatial correction term),
column~2 gives the weighted mean and error of the offset between the observed GEDR3 parallax and the photometric parallax,
column~3 gives the weighted mean and error of the L20 correction,
columns~4-7 give the weighted mean and error for the spatial correction, the magnitude correction, the total correction,
and the offset between the corrected GEDR3 parallax in the present work (PW) and the photometric parallax.
Column~8 gives the number of stars.
The sample sizes of 66 and 54 stars are explained in Sect.~\ref{SS:CC}.
\tablefoottext{a}{This model results in a $\Delta$(PZPO corrected parallax) of $0.00 \pm 2.93$~\muas\ when using the L20 correction.}
}
\label{Table:CC}
\end{table*}

\section{Discussion and summary}
\label{S:Disc}

The presence of a parallax zero point offset that was identified in GDR2 received a lot of attention.
The L20 paper analysing the new GEDR3 data offers a lot of insight into the issue and they presented a python script to calculate
the correction based on $G$, $\beta$, and the pseudocolour or $\nu_{\rm eff}$ (depending on the {\tt astrometric\_params\_solved} parameter).

On the other hand, L20 remark that `the results should $\ldots$ in no way be regarded as definitive', and that
`alternative routes are explored towards getting a better handle on the systematics in Gaia data'. 
The present paper should be viewed in this light. An alternative procedure to the one in L20 is proposed which is offered
to the community for further scrutiny.

The two approaches are similar in that both use samples of QSOs and wide binaries (albeit selected in different ways).
The main differences to the L20 approach are that (1) there is no separation between 5- and 6-parameter solutions,
(2) the colour dependence uses the $(G_\text{BP} - G_\text{RP})$ colour rather than pseudocolour or $\nu_{\rm eff}$, 
(3) the dependence on sky position and magnitude are separated, and are treated as additive terms, and that
(4) the present approach gives a correction including an error estimate.

It is shown that the PZPO shows a more complicated behaviour than only on the ecliptic latitude (Fig.~\ref{Fig:QSOparoff}, also
see \citealt{Huang21}).
L20 argue that such a dependence is theoretically expected and related to the scanning law but this would not explain
the different behaviour at bright (Fig.~\ref{Fig:CCs}; a range of $\sim$35~\muas\ with the largest correction around $\beta \sim 5$\degr)
and faint magnitudes (Fig.~\ref{Fig:Correction}; a range of $\sim$20~\muas\ with a slow increase with $\beta$) in the L20 recipe.

Here, the practical approach is taken to calculate the PZPO over the sky using the HEALPix formalism.
Using the dependence of the PZPO as a function of $G$, a spatial PZPO at $G$ = 20~mag is determined for several HEALPix levels, based on the
QSO sample for $G >$ 17~mag.
A large sample of WBs with very low chance alignments is used to derive the magnitude dependence of the PZPO for magnitudes $<$19~mag.
The range of $17 - 19$~mag is used to connect the QSO to the WB sample.

The L20 recipe does not provide an error on the correction. It is shown here that error on the PZPO is dominated
by the error on the spatial correction, and that it can be substantial (up to several tens of \muas\ depending on sky position).
Increasing the sample of QSOs, especially in the direction of the Galactic plane, will help in reducing the statistical error but
only as $1/\sqrt{N}$. 

The recipes provided here can not be easily transformed into a simple script. This may be seen as a disadvantage, on the other hand
is requires the user to make informed choices.
The procedure to be followed is as follows:
  \begin{itemize}
  \item Obtain the {\tt source\_id}, $G$ magnitude, parallax and error, and $(G_\text{BP} - G_\text{RP})$ colour from GEDR3 for your source(s).
  \item Get the pixel number in the HEALPix scheme from the {\tt source\_id} following footnote~\ref{FN2} for levels 0, 1, 2, 3, and 4.
  \item Use the results from models 30-34 available from the CDS to obtain the spatial correction and error at $G$ = 20 for the various HEALPix levels.
  \item For $G >$19.9~mag apply Eq.~\ref{Eq-cor}, otherwise apply Eq.~\ref{Eq-corWB}.
    The error in this correction is a 1~\muas\ systematic error to be added in quadrature to a random error of 2.7~\muas\ for $G > 6$
    and 13~\muas\ for $G \le 6$.
  \item If a colour term is to be included use the results from models 35-39, and additionally apply Eq.~\ref{Eq-cor1}.
    This colour term is derived for the QSOs sample ($G$ \more 17~mag, 0.2~\less~$(G_\text{BP}-G_\text{RP}$)~\less~1.6~mag) and is untested
    outside this range.
  \item Add the spatial and magnitude (+colour) correction, and add the errors in quadrature. Subtract the total from the observed parallax to
    obtain the corrected parallax, that is, an estimate of the true parallax (Eq.~\ref{Eq-basic}).
    Also in this last step the errors should be added in quadrature.

  \end{itemize}

Following the examples described in Section~\ref{S-Appl}, it is recommended
do this for all available HEALPix levels and then choose the highest level that
does not compromise the S/N.

\begin{acknowledgements}
This work has made use of data from the European Space Agency (ESA) mission {\it Gaia}
(\url{https://www.cosmos.esa.int/gaia}), processed by the {\it Gaia} Data Processing and Analysis Consortium
(DPAC, \url{https://www.cosmos.esa.int/web/gaia/dpac/consortium}). Funding for the DPAC
has been provided by national institutions, in particular the institutions
participating in the {\it Gaia} Multilateral Agreement.
This research has made use of the SIMBAD database, the VizieR catalogue access tool and
the cross-match service provided by CDS, Strasbourg.
Thanks to Francois-Xavier Pineau for explaining the best use of the cross-match service.
Some of the results in this paper have been derived using the healpy and HEALPix package.
I would like to thank Drs. Kareen El-Badry and Valentin Ivanov for discussions on
wide binaries and quasar catalogues, respectively, and the referee for a careful reading of the manuscript and helpful suggestions.

\end{acknowledgements}

\bibliographystyle{aa.bst}
\bibliography{references.bib}

\begin{thebibliography}{45}
\expandafter\ifx\csname natexlab\endcsname\relax\def\natexlab#1{#1}\fi

\bibitem[{{Anderson}(2018)}]{Anderson18}
{Anderson}, R.~I. 2018, \aap, 611, L7

\bibitem[{{Benedict} {et~al.}(2017){Benedict}, {McArthur}, {Nelan}, \&
  {Harrison}}]{Benedict17}
{Benedict}, G.~F., {McArthur}, B.~E., {Nelan}, E.~P., \& {Harrison}, T.~E.
  2017, \pasp, 129, 012001

\bibitem[{{Bhardwaj} {et~al.}(2021){Bhardwaj}, {Rejkuba}, {de Grijs}, {Yang},
  {Herczeg}, {Marconi}, {Singh}, {Kanbur}, \& {Ngeow}}]{Bhardwaj20}
{Bhardwaj}, A., {Rejkuba}, M., {de Grijs}, R., {et~al.} 2021, \apj, 909, 200

\bibitem[{{Bond} {et~al.}(2018){Bond}, {Nelan}, {Remage Evans}, {Schaefer}, \&
  {Harmer}}]{Bond18}
{Bond}, H.~E., {Nelan}, E.~P., {Remage Evans}, N., {Schaefer}, G.~H., \&
  {Harmer}, D. 2018, \apj, 853, 55

\bibitem[{{Casertano} {et~al.}(2016){Casertano}, {Riess}, {Anderson},
  {Anderson}, {Bowers}, {Clubb}, {Cukierman}, {Filippenko}, {Graham},
  {MacKenty}, {Melis}, {Tucker}, \& {Upadhya}}]{Casertano16}
{Casertano}, S., {Riess}, A.~G., {Anderson}, J., {et~al.} 2016, \apj, 825, 11

\bibitem[{{Chan} \& {Bovy}(2020)}]{ChanBovy19}
{Chan}, V.~C. \& {Bovy}, J. 2020, \mnras, 493, 4367

\bibitem[{{Clementini} {et~al.}(2019){Clementini}, {Ripepi}, {Molinaro},
  {Garofalo}, {Muraveva}, {Rimoldini}, {Guy}, {Jevardat de Fombelle},
  {Nienartowicz}, {Marchal}, {Audard}, {Holl}, {Leccia}, {Marconi}, {Musella},
  {Mowlavi}, {Lecoeur-Taibi}, {Eyer}, {De Ridder}, {Regibo}, {Sarro},
  {Szabados}, {Evans}, \& {Riello}}]{Clementini19}
{Clementini}, G., {Ripepi}, V., {Molinaro}, R., {et~al.} 2019, \aap, 622, A60

\bibitem[{{El-Badry} {et~al.}(2021){El-Badry}, {Rix}, \& {Heintz}}]{ElBadry21}
{El-Badry}, K., {Rix}, H.-W., \& {Heintz}, T.~M. 2021, \mnras
  [\eprint[arXiv]{2101.05282}]

\bibitem[{{Fabricius} {et~al.}(2021){Fabricius}, {Luri}, {Arenou}, {Babusiaux},
  {Helmi}, {Muraveva}, {Reyl{\'e}}, {Spoto}, {Vallenari}, {Antoja}, {Balbinot},
  {Barache}, {Bauchet}, {Bragaglia}, {Busonero}, {Cantat-Gaudin}, {Carrasco},
  {Diakit{\'e}}, {Fabrizio}, {Figueras}, {Garcia-Gutierrez}, {Garofalo},
  {Jordi}, {Kervella}, {Khanna}, {Leclerc}, {Licata}, {Lambert}, {Marrese},
  {Masip}, {Ramos}, {Robichon}, {Robin}, {Romero-G{\'o}mez}, {Rubele}, \&
  {Weiler}}]{Fabricius20}
{Fabricius}, C., {Luri}, X., {Arenou}, F., {et~al.} 2021, \aap, 649, A5

\bibitem[{{Flesch}(2019)}]{Flesch19}
{Flesch}, E.~W. 2019, arXiv e-prints, arXiv:1912.05614

\bibitem[{{Gaia Collaboration} {et~al.}(2018){Gaia Collaboration}, {Brown},
  {Vallenari}, {Prusti}, {de Bruijne}, {Babusiaux}, {Bailer-Jones}, {Biermann},
  {Evans}, {Eyer}, \& et~al.}]{GDR2Sum}
{Gaia Collaboration}, {Brown}, A.~G.~A., {Vallenari}, A., {et~al.} 2018, \aap,
  616, A1

\bibitem[{{Gaia Collaboration} {et~al.}(2021{\natexlab{a}}){Gaia
  Collaboration}, {Brown}, {Vallenari}, {Prusti}, {de Bruijne}, {Babusiaux},
  {Biermann}, {Creevey}, {Evans}, {Eyer}, {Hutton}, {Jansen}, {Jordi},
  {Klioner}, {Lammers}, {Lindegren}, {Luri}, {Mignard}, {Panem}, {Pourbaix},
  {Randich}, {Sartoretti}, {Soubiran}, {Walton}, {Arenou}, {Bailer-Jones},
  {Bastian}, {Cropper}, {Drimmel}, {Katz}, {Lattanzi}, {van Leeuwen}, {Bakker},
  {Cacciari}, {Casta{\~n}eda}, {De Angeli}, {Ducourant}, {Fabricius},
  {Fouesneau}, {Fr{\'e}mat}, {Guerra}, {Guerrier}, {Guiraud}, {Jean-Antoine
  Piccolo}, {Masana}, {Messineo}, {Mowlavi}, {Nicolas}, {Nienartowicz},
  {Pailler}, {Panuzzo}, {Riclet}, {Roux}, {Seabroke}, {Sordo}, {Tanga},
  {Th{\'e}venin}, {Gracia-Abril}, {Portell}, {Teyssier}, {Altmann}, {Andrae},
  {Bellas-Velidis}, {Benson}, {Berthier}, {Blomme}, {Brugaletta}, {Burgess},
  {Busso}, {Carry}, {Cellino}, {Cheek}, {Clementini}, {Damerdji}, {Davidson},
  {Delchambre}, {Dell'Oro}, {Fern{\'a}ndez-Hern{\'a}ndez}, {Galluccio},
  {Garc{\'\i}a-Lario}, {Garcia-Reinaldos}, {Gonz{\'a}lez-N{\'u}{\~n}ez},
  {Gosset}, {Haigron}, {Halbwachs}, {Hambly}, {Harrison}, {Hatzidimitriou},
  {Heiter}, {Hern{\'a}ndez}, {Hestroffer}, {Hodgkin}, {Holl}, {Jan{\ss}en},
  {Jevardat de Fombelle}, {Jordan}, {Krone-Martins}, {Lanzafame},
  {L{\"o}ffler}, {Lorca}, {Manteiga}, {Marchal}, {Marrese}, {Moitinho}, {Mora},
  {Muinonen}, {Osborne}, {Pancino}, {Pauwels}, {Petit}, {Recio-Blanco},
  {Richards}, {Riello}, {Rimoldini}, {Robin}, {Roegiers}, {Rybizki}, {Sarro},
  {Siopis}, {Smith}, {Sozzetti}, {Ulla}, {Utrilla}, {van Leeuwen}, {van
  Reeven}, {Abbas}, {Abreu Aramburu}, {Accart}, {Aerts}, {Aguado}, {Ajaj},
  {Altavilla}, {{\'A}lvarez}, {{\'A}lvarez Cid-Fuentes}, {Alves}, {Anderson},
  {Anglada Varela}, {Antoja}, {Audard}, {Baines}, {Baker},
  {Balaguer-N{\'u}{\~n}ez}, {Balbinot}, {Balog}, {Barache}, {Barbato},
  {Barros}, {Barstow}, {Bartolom{\'e}}, {Bassilana}, {Bauchet},
  {Baudesson-Stella}, {Becciani}, {Bellazzini}, {Bernet}, {Bertone}, {Bianchi},
  {Blanco-Cuaresma}, {Boch}, {Bombrun}, {Bossini}, {Bouquillon}, {Bragaglia},
  {Bramante}, {Breedt}, {Bressan}, {Brouillet}, {Bucciarelli}, {Burlacu},
  {Busonero}, {Butkevich}, {Buzzi}, {Caffau}, {Cancelliere}, {C{\'a}novas},
  {Cantat-Gaudin}, {Carballo}, {Carlucci}, {Carnerero}, {Carrasco},
  {Casamiquela}, {Castellani}, {Castro-Ginard}, {Castro Sampol}, {Chaoul},
  {Charlot}, {Chemin}, {Chiavassa}, {Cioni}, {Comoretto}, {Cooper}, {Cornez},
  {Cowell}, {Crifo}, {Crosta}, {Crowley}, {Dafonte}, {Dapergolas}, {David},
  {David}, {de Laverny}, {De Luise}, {De March}, {De Ridder}, {de Souza}, {de
  Teodoro}, {de Torres}, {del Peloso}, {del Pozo}, {Delbo}, {Delgado},
  {Delgado}, {Delisle}, {Di Matteo}, {Diakite}, {Diener}, {Distefano},
  {Dolding}, {Eappachen}, {Edvardsson}, {Enke}, {Esquej}, {Fabre}, {Fabrizio},
  {Faigler}, {Fedorets}, {Fernique}, {Fienga}, {Figueras}, {Fouron},
  {Fragkoudi}, {Fraile}, {Franke}, {Gai}, {Garabato}, {Garcia-Gutierrez},
  {Garc{\'\i}a-Torres}, {Garofalo}, {Gavras}, {Gerlach}, {Geyer}, {Giacobbe},
  {Gilmore}, {Girona}, {Giuffrida}, {Gomel}, {Gomez}, {Gonzalez-Santamaria},
  {Gonz{\'a}lez-Vidal}, {Granvik}, {Guti{\'e}rrez-S{\'a}nchez}, {Guy},
  {Hauser}, {Haywood}, {Helmi}, {Hidalgo}, {Hilger}, {H{\l}adczuk}, {Hobbs},
  {Holland}, {Huckle}, {Jasniewicz}, {Jonker}, {Juaristi Campillo}, {Julbe},
  {Karbevska}, {Kervella}, {Khanna}, {Kochoska}, {Kontizas}, {Kordopatis},
  {Korn}, {Kostrzewa-Rutkowska}, {Kruszy{\'n}ska}, {Lambert}, {Lanza}, {Lasne},
  {Le Campion}, {Le Fustec}, {Lebreton}, {Lebzelter}, {Leccia}, {Leclerc},
  {Lecoeur-Taibi}, {Liao}, {Licata}, {Lindstr{\o}m}, {Lister}, {Livanou},
  {Lobel}, {Madrero Pardo}, {Managau}, {Mann}, {Marchant}, {Marconi}, {Marcos
  Santos}, {Marinoni}, {Marocco}, {Marshall}, {Martin Polo},
  {Mart{\'\i}n-Fleitas}, {Masip}, {Massari}, {Mastrobuono-Battisti}, {Mazeh},
  {McMillan}, {Messina}, {Michalik}, {Millar}, {Mints}, {Molina}, {Molinaro},
  {Moln{\'a}r}, {Montegriffo}, {Mor}, {Morbidelli}, {Morel}, {Morris},
  {Mulone}, {Munoz}, {Muraveva}, {Murphy}, {Musella}, {Noval}, {Ord{\'e}novic},
  {Orr{\`u}}, {Osinde}, {Pagani}, {Pagano}, {Palaversa}, {Palicio}, {Panahi},
  {Pawlak}, {Pe{\~n}alosa Esteller}, {Penttil{\"a}}, {Piersimoni}, {Pineau},
  {Plachy}, {Plum}, {Poggio}, {Poretti}, {Poujoulet}, {Pr{\v{s}}a}, {Pulone},
  {Racero}, {Ragaini}, {Rainer}, {Raiteri}, {Rambaux}, {Ramos}, {Ramos-Lerate},
  {Re Fiorentin}, {Regibo}, {Reyl{\'e}}, {Ripepi}, {Riva}, {Rixon}, {Robichon},
  {Robin}, {Roelens}, {Rohrbasser}, {Romero-G{\'o}mez}, {Rowell}, {Royer},
  {Rybicki}, {Sadowski}, {Sagrist{\`a} Sell{\'e}s}, {Sahlmann}, {Salgado},
  {Salguero}, {Samaras}, {Sanchez Gimenez}, {Sanna}, {Santove{\~n}a},
  {Sarasso}, {Schultheis}, {Sciacca}, {Segol}, {Segovia}, {S{\'e}gransan},
  {Semeux}, {Shahaf}, {Siddiqui}, {Siebert}, {Siltala}, {Slezak}, {Smart},
  {Solano}, {Solitro}, {Souami}, {Souchay}, {Spagna}, {Spoto}, {Steele},
  {Steidelm{\"u}ller}, {Stephenson}, {S{\"u}veges}, {Szabados}, {Szegedi-Elek},
  {Taris}, {Tauran}, {Taylor}, {Teixeira}, {Thuillot}, {Tonello}, {Torra},
  {Torra}, {Turon}, {Unger}, {Vaillant}, {van Dillen}, {Vanel}, {Vecchiato},
  {Viala}, {Vicente}, {Voutsinas}, {Weiler}, {Wevers}, {Wyrzykowski}, {Yoldas},
  {Yvard}, {Zhao}, {Zorec}, {Zucker}, {Zurbach}, \& {Zwitter}}]{GEDR3_Brown}
{Gaia Collaboration}, {Brown}, A.~G.~A., {Vallenari}, A., {et~al.}
  2021{\natexlab{a}}, \aap, 649, A1

\bibitem[{{Gaia Collaboration} {et~al.}(2021{\natexlab{b}}){Gaia
  Collaboration}, {Klioner}, {Mignard}, {Lindegren}, {Bastian}, {McMillan},
  {Hern{\'a}ndez}, {Hobbs}, {Ramos-Lerate}, {Biermann}, {Bombrun}, {de Torres},
  {Gerlach}, {Geyer}, {Hilger}, {Lammers}, {Steidelm{\"u}ller}, {Stephenson},
  {Brown}, {Vallenari}, {Prusti}, {de Bruijne}, {Babusiaux}, {Creevey},
  {Evans}, {Eyer}, {Hutton}, {Jansen}, {Jordi}, {Luri}, {Panem}, {Pourbaix},
  {Randich}, {Sartoretti}, {Soubiran}, {Walton}, {Arenou}, {Bailer-Jones},
  {Cropper}, {Drimmel}, {Katz}, {Lattanzi}, {van Leeuwen}, {Bakker},
  {Casta{\~n}eda}, {De Angeli}, {Ducourant}, {Fabricius}, {Fouesneau},
  {Fr{\'e}mat}, {Guerra}, {Guerrier}, {Guiraud}, {Jean-Antoine Piccolo},
  {Masana}, {Messineo}, {Mowlavi}, {Nicolas}, {Nienartowicz}, {Pailler},
  {Panuzzo}, {Riclet}, {Roux}, {Seabroke}, {Sordo}, {Tanga}, {Th{\'e}venin},
  {Gracia-Abril}, {Portell}, {Teyssier}, {Altmann}, {Andrae}, {Bellas-Velidis},
  {Benson}, {Berthier}, {Blomme}, {Brugaletta}, {Burgess}, {Busso}, {Carry},
  {Cellino}, {Cheek}, {Clementini}, {Damerdji}, {Davidson}, {Delchambre},
  {Dell'Oro}, {Fern{\'a}ndez-Hern{\'a}ndez}, {Galluccio}, {Garc{\'\i}a-Lario},
  {Garcia-Reinaldos}, {Gonz{\'a}lez-N{\'u}{\~n}ez}, {Gosset}, {Haigron},
  {Halbwachs}, {Hambly}, {Harrison}, {Hatzidimitriou}, {Heiter}, {Hestroffer},
  {Hodgkin}, {Holl}, {Jan{\ss}en}, {Jevardat de Fombelle}, {Jordan},
  {Krone-Martins}, {Lanzafame}, {L{\"o}ffler}, {Lorca}, {Manteiga}, {Marchal},
  {Marrese}, {Moitinho}, {Mora}, {Muinonen}, {Osborne}, {Pancino}, {Pauwels},
  {Recio-Blanco}, {Richards}, {Riello}, {Rimoldini}, {Robin}, {Roegiers},
  {Rybizki}, {Sarro}, {Siopis}, {Smith}, {Sozzetti}, {Ulla}, {Utrilla}, {van
  Leeuwen}, {van Reeven}, {Abbas}, {Abreu Aramburu}, {Accart}, {Aerts},
  {Aguado}, {Ajaj}, {Altavilla}, {{\'A}lvarez}, {{\'A}lvarez Cid-Fuentes},
  {Alves}, {Anderson}, {Anglada Varela}, {Antoja}, {Audard}, {Baines}, {Baker},
  {Balaguer-N{\'u}{\~n}ez}, {Balbinot}, {Balog}, {Barache}, {Barbato},
  {Barros}, {Barstow}, {Bartolom{\'e}}, {Bassilana}, {Bauchet},
  {Baudesson-Stella}, {Becciani}, {Bellazzini}, {Bernet}, {Bertone}, {Bianchi},
  {Blanco-Cuaresma}, {Boch}, {Bossini}, {Bouquillon}, {Bramante}, {Breedt},
  {Bressan}, {Brouillet}, {Bucciarelli}, {Burlacu}, {Busonero}, {Butkevich},
  {Buzzi}, {Caffau}, {Cancelliere}, {C{\'a}novas}, {Cantat-Gaudin}, {Carballo},
  {Carlucci}, {Carnerero}, {Carrasco}, {Casamiquela}, {Castellani},
  {Castro-Ginard}, {Castro Sampol}, {Chaoul}, {Charlot}, {Chemin}, {Chiavassa},
  {Comoretto}, {Cooper}, {Cornez}, {Cowell}, {Crifo}, {Crosta}, {Crowley},
  {Dafonte}, {Dapergolas}, {David}, {David}, {de Laverny}, {De Luise}, {De
  March}, {De Ridder}, {de Souza}, {de Teodoro}, {del Peloso}, {del Pozo},
  {Delgado}, {Delgado}, {Delisle}, {Di Matteo}, {Diakite}, {Diener},
  {Distefano}, {Dolding}, {Eappachen}, {Enke}, {Esquej}, {Fabre}, {Fabrizio},
  {Faigler}, {Fedorets}, {Fernique}, {Fienga}, {Figueras}, {Fouron},
  {Fragkoudi}, {Fraile}, {Franke}, {Gai}, {Garabato}, {Garcia-Gutierrez},
  {Garc{\'\i}a-Torres}, {Garofalo}, {Gavras}, {Giacobbe}, {Gilmore}, {Girona},
  {Giuffrida}, {Gomez}, {Gonzalez-Santamaria}, {Gonz{\'a}lez-Vidal}, {Granvik},
  {Guti{\'e}rrez-S{\'a}nchez}, {Guy}, {Hauser}, {Haywood}, {Helmi}, {Hidalgo},
  {H{\l}adczuk}, {Holland}, {Huckle}, {Jasniewicz}, {Jonker}, {Juaristi
  Campillo}, {Julbe}, {Karbevska}, {Kervella}, {Khanna}, {Kochoska},
  {Kordopatis}, {Korn}, {Kostrzewa-Rutkowska}, {Kruszy{\'n}ska}, {Lambert},
  {Lanza}, {Lasne}, {Le Campion}, {Le Fustec}, {Lebreton}, {Lebzelter},
  {Leccia}, {Leclerc}, {Lecoeur-Taibi}, {Liao}, {Licata}, {Lindstr{\o}m},
  {Lister}, {Livanou}, {Lobel}, {Madrero Pardo}, {Managau}, {Mann}, {Marchant},
  {Marconi}, {Marcos Santos}, {Marinoni}, {Marocco}, {Marshall}, {Martin Polo},
  {Mart{\'\i}n-Fleitas}, {Masip}, {Massari}, {Mastrobuono-Battisti}, {Mazeh},
  {Messina}, {Michalik}, {Millar}, {Mints}, {Molina}, {Molinaro}, {Moln{\'a}r},
  {Montegriffo}, {Mor}, {Morbidelli}, {Morel}, {Morris}, {Mulone}, {Munoz},
  {Muraveva}, {Murphy}, {Musella}, {Noval}, {Ord{\'e}novic}, {Orr{\`u}},
  {Osinde}, {Pagani}, {Pagano}, {Palaversa}, {Palicio}, {Panahi}, {Pawlak},
  {Pe{\~n}alosa Esteller}, {Penttil{\"a}}, {Piersimoni}, {Pineau}, {Plachy},
  {Plum}, {Poggio}, {Poretti}, {Poujoulet}, {Pr{\v{s}}a}, {Pulone}, {Racero},
  {Ragaini}, {Rainer}, {Raiteri}, {Rambaux}, {Ramos}, {Re Fiorentin}, {Regibo},
  {Reyl{\'e}}, {Ripepi}, {Riva}, {Rixon}, {Robichon}, {Robin}, {Roelens},
  {Rohrbasser}, {Romero-G{\'o}mez}, {Rowell}, {Royer}, {Rybicki}, {Sadowski},
  {Sagrist{\`a} Sell{\'e}s}, {Sahlmann}, {Salgado}, {Salguero}, {Samaras},
  {Sanchez Gimenez}, {Sanna}, {Santove{\~n}a}, {Sarasso}, {Schultheis},
  {Sciacca}, {Segol}, {Segovia}, {S{\'e}gransan}, {Semeux}, {Siddiqui},
  {Siebert}, {Siltala}, {Slezak}, {Smart}, {Solano}, {Solitro}, {Souami},
  {Souchay}, {Spagna}, {Spoto}, {Steele}, {S{\"u}veges}, {Szabados},
  {Szegedi-Elek}, {Taris}, {Tauran}, {Taylor}, {Teixeira}, {Thuillot},
  {Tonello}, {Torra}, {Torra}, {Turon}, {Unger}, {Vaillant}, {van Dillen},
  {Vanel}, {Vecchiato}, {Viala}, {Vicente}, {Voutsinas}, {Weiler}, {Wevers},
  {Wyrzykowski}, {Yoldas}, {Yvard}, {Zhao}, {Zorec}, {Zucker}, {Zurbach}, \&
  {Zwitter}}]{GEDR3_Klioner}
{Gaia Collaboration}, {Klioner}, S.~A., {Mignard}, F., {et~al.}
  2021{\natexlab{b}}, \aap, 649, A9

\bibitem[{{Gaia Collaboration} {et~al.}(2016){Gaia Collaboration}, {Prusti},
  {de Bruijne}, {Brown}, {Vallenari}, {Babusiaux}, {Bailer-Jones}, {Bastian},
  {Biermann}, {Evans}, \& et~al.}]{GC2016a}
{Gaia Collaboration}, {Prusti}, T., {de Bruijne}, J.~H.~J., {et~al.} 2016,
  \aap, 595, A1

\bibitem[{{G{\'o}rski} {et~al.}(2005){G{\'o}rski}, {Hivon}, {Banday},
  {Wandelt}, {Hansen}, {Reinecke}, \& {Bartelmann}}]{HEALPix05}
{G{\'o}rski}, K.~M., {Hivon}, E., {Banday}, A.~J., {et~al.} 2005, \apj, 622,
  759

\bibitem[{{Graczyk} {et~al.}(2019){Graczyk}, {Pietrzy{\'n}ski}, {Gieren},
  {Storm}, {Nardetto}, {Gallenne}, {Maxted}, {Kervella}, {Ko{\l}aczkowski},
  {Konorski}, {Pilecki}, {Zgirski}, {G{\'o}rski}, {Suchomska}, {Karczmarek},
  {Taormina}, {Wielg{\'o}rski}, {Narloch}, {Smolec}, {Chini}, \&
  {Breuval}}]{Graczyk19}
{Graczyk}, D., {Pietrzy{\'n}ski}, G., {Gieren}, W., {et~al.} 2019, \apj, 872,
  85

\bibitem[{{Groenewegen}(2018)}]{Gr_GDR2}
{Groenewegen}, M.~A.~T. 2018, \aap, 619, A8

\bibitem[{{Huang} {et~al.}(2020){Huang}, {Sch{\"o}nrich}, {Zhang}, {Wu},
  {Chen}, {Wang}, {Xiang}, {Wang}, {Yuan}, {Li}, {Sun}, {Li}, \&
  {Liu}}]{Huang20}
{Huang}, Y., {Sch{\"o}nrich}, R., {Zhang}, H., {et~al.} 2020, \apjs, 249, 29

\bibitem[{{Huang} {et~al.}(2021){Huang}, {Yuan}, {Beers}, \& {Zhang}}]{Huang21}
{Huang}, Y., {Yuan}, H., {Beers}, T.~C., \& {Zhang}, H. 2021, \apjl, 910, L5

\bibitem[{{Khan} {et~al.}(2019){Khan}, {Miglio}, {Mosser}, {Arenou},
  {Belkacem}, {Brown}, {Katz}, {Casagrande}, {Chaplin}, {Davies}, {Rendle},
  {Rodrigues}, {Bossini}, {Cantat-Gaudin}, {Elsworth}, {Girardi}, {North}, \&
  {Vallenari}}]{Khan19}
{Khan}, S., {Miglio}, A., {Mosser}, B., {et~al.} 2019, in The Gaia Universe, 13

\bibitem[{{Layden} {et~al.}(2019){Layden}, {Tiede}, {Chaboyer}, {Bunner}, \&
  {Smitka}}]{Layden19}
{Layden}, A.~C., {Tiede}, G.~P., {Chaboyer}, B., {Bunner}, C., \& {Smitka},
  M.~T. 2019, \aj, 158, 105

\bibitem[{{Leung} \& {Bovy}(2019)}]{LeungBovy19}
{Leung}, H.~W. \& {Bovy}, J. 2019, \mnras, 489, 2079

\bibitem[{Lindegren(2018)}]{LindegrenTN}
Lindegren, L. 2018, {G}AIA-C3-TN-LU-LL-124

\bibitem[{{Lindegren} {et~al.}(2021{\natexlab{a}}){Lindegren}, {Bastian},
  {Biermann}, {Bombrun}, {de Torres}, {Gerlach}, {Geyer}, {Hern{\'a}ndez},
  {Hilger}, {Hobbs}, {Klioner}, {Lammers}, {McMillan}, {Ramos-Lerate},
  {Steidelm{\"u}ller}, {Stephenson}, \& {van Leeuwen}}]{GEDR3_LindegrenZP}
{Lindegren}, L., {Bastian}, U., {Biermann}, M., {et~al.} 2021{\natexlab{a}},
  \aap, 649, A4

\bibitem[{{Lindegren} {et~al.}(2018){Lindegren}, {Hern{\'a}ndez}, {Bombrun},
  {Klioner}, {Bastian}, {Ramos-Lerate}, {de Torres}, {Steidelm{\"u}ller},
  {Stephenson}, {Hobbs}, {Lammers}, {Biermann}, {Geyer}, {Hilger}, {Michalik},
  {Stampa}, {McMillan}, {Casta{\~n}eda}, {Clotet}, {Comoretto}, {Davidson},
  {Fabricius}, {Gracia}, {Hambly}, {Hutton}, {Mora}, {Portell}, {van Leeuwen},
  {Abbas}, {Abreu}, {Altmann}, {Andrei}, {Anglada}, {Balaguer-N{\'u}{\~n}ez},
  {Barache}, {Becciani}, {Bertone}, {Bianchi}, {Bouquillon}, {Bourda},
  {Br{\"u}semeister}, {Bucciarelli}, {Busonero}, {Buzzi}, {Cancelliere},
  {Carlucci}, {Charlot}, {Cheek}, {Crosta}, {Crowley}, {de Bruijne}, {de
  Felice}, {Drimmel}, {Esquej}, {Fienga}, {Fraile}, {Gai}, {Garralda},
  {Gonz{\'a}lez-Vidal}, {Guerra}, {Hauser}, {Hofmann}, {Holl}, {Jordan},
  {Lattanzi}, {Lenhardt}, {Liao}, {Licata}, {Lister}, {L{\"o}ffler},
  {Marchant}, {Martin-Fleitas}, {Messineo}, {Mignard}, {Morbidelli}, {Poggio},
  {Riva}, {Rowell}, {Salguero}, {Sarasso}, {Sciacca}, {Siddiqui}, {Smart},
  {Spagna}, {Steele}, {Taris}, {Torra}, {van Elteren}, {van Reeven}, \&
  {Vecchiato}}]{Lindegren18}
{Lindegren}, L., {Hern{\'a}ndez}, J., {Bombrun}, A., {et~al.} 2018, \aap, 616,
  A2

\bibitem[{{Lindegren} {et~al.}(2021{\natexlab{b}}){Lindegren}, {Klioner},
  {Hern{\'a}ndez}, {Bombrun}, {Ramos-Lerate}, {Steidelm{\"u}ller}, {Bastian},
  {Biermann}, {de Torres}, {Gerlach}, {Geyer}, {Hilger}, {Hobbs}, {Lammers},
  {McMillan}, {Stephenson}, {Casta{\~n}eda}, {Davidson}, {Fabricius},
  {Gracia-Abril}, {Portell}, {Rowell}, {Teyssier}, {Torra}, {Bartolom{\'e}},
  {Clotet}, {Garralda}, {Gonz{\'a}lez-Vidal}, {Torra}, {Abbas}, {Altmann},
  {Anglada Varela}, {Balaguer-N{\'u}{\~n}ez}, {Balog}, {Barache}, {Becciani},
  {Bernet}, {Bertone}, {Bianchi}, {Bouquillon}, {Brown}, {Bucciarelli},
  {Busonero}, {Butkevich}, {Buzzi}, {Cancelliere}, {Carlucci}, {Charlot},
  {Cioni}, {Crosta}, {Crowley}, {del Peloso}, {del Pozo}, {Drimmel}, {Esquej},
  {Fienga}, {Fraile}, {Gai}, {Garcia-Reinaldos}, {Guerra}, {Hambly}, {Hauser},
  {Jan{\ss}en}, {Jordan}, {Kostrzewa-Rutkowska}, {Lattanzi}, {Liao}, {Licata},
  {Lister}, {L{\"o}ffler}, {Marchant}, {Masip}, {Mignard}, {Mints}, {Molina},
  {Mora}, {Morbidelli}, {Murphy}, {Pagani}, {Panuzzo}, {Pe{\~n}alosa Esteller},
  {Poggio}, {Re Fiorentin}, {Riva}, {Sagrist{\`a} Sell{\'e}s}, {Sanchez
  Gimenez}, {Sarasso}, {Sciacca}, {Siddiqui}, {Smart}, {Souami}, {Spagna},
  {Steele}, {Taris}, {Utrilla}, {van Reeven}, \&
  {Vecchiato}}]{GEDR3_LindegrenAS}
{Lindegren}, L., {Klioner}, S.~A., {Hern{\'a}ndez}, J., {et~al.}
  2021{\natexlab{b}}, \aap, 649, A2

\bibitem[{{Muraveva} {et~al.}(2018){Muraveva}, {Delgado}, {Clementini},
  {Sarro}, \& {Garofalo}}]{Muraveva18}
{Muraveva}, T., {Delgado}, H.~E., {Clementini}, G., {Sarro}, L.~M., \&
  {Garofalo}, A. 2018, \mnras, 481, 1195

\bibitem[{Press {et~al.}(1992)Press, Teukolsky, Vetterling, \&
  Flannery}]{Press1992}
Press, W., Teukolsky, S., Vetterling, W., \& Flannery, B. 1992, {Numerical
  Recipes in C} (Cambridge: Cambridge University Press)

\bibitem[{{Ren} {et~al.}(2021){Ren}, {Chen}, {Zhang}, {de Grijs}, {Deng}, \&
  {Huang}}]{Ren21}
{Ren}, F., {Chen}, X., {Zhang}, H., {et~al.} 2021, \apjl, 911, L20

\bibitem[{{Riess} {et~al.}(2014){Riess}, {Casertano}, {Anderson}, {MacKenty},
  \& {Filippenko}}]{Riess14}
{Riess}, A.~G., {Casertano}, S., {Anderson}, J., {MacKenty}, J., \&
  {Filippenko}, A.~V. 2014, \apj, 785, 161

\bibitem[{{Riess} {et~al.}(2021){Riess}, {Casertano}, {Yuan}, {Bowers},
  {Macri}, {Zinn}, \& {Scolnic}}]{RiessGEDR3}
{Riess}, A.~G., {Casertano}, S., {Yuan}, W., {et~al.} 2021, \apjl, 908, L6

\bibitem[{{Riess} {et~al.}(2018{\natexlab{a}}){Riess}, {Casertano}, {Yuan},
  {Macri}, {Anderson}, {MacKenty}, {Bowers}, {Clubb}, {Filippenko}, {Jones}, \&
  {Tucker}}]{Riess18}
{Riess}, A.~G., {Casertano}, S., {Yuan}, W., {et~al.} 2018{\natexlab{a}}, \apj,
  855, 136

\bibitem[{{Riess} {et~al.}(2018{\natexlab{b}}){Riess}, {Casertano}, {Yuan},
  {Macri}, {Bucciarelli}, {Lattanzi}, {MacKenty}, {Bowers}, {Zheng},
  {Filippenko}, {Huang}, \& {Anderson}}]{RiessGDR2}
{Riess}, A.~G., {Casertano}, S., {Yuan}, W., {et~al.} 2018{\natexlab{b}}, \apj,
  861, 126

\bibitem[{{Riess} {et~al.}(2019){Riess}, {Casertano}, {Yuan}, {Macri}, \&
  {Scolnic}}]{Riess19}
{Riess}, A.~G., {Casertano}, S., {Yuan}, W., {Macri}, L.~M., \& {Scolnic}, D.
  2019, \apj, 876, 85

\bibitem[{{Riess} {et~al.}(2016){Riess}, {Macri}, {Hoffmann}, {Scolnic},
  {Casertano}, {Filippenko}, {Tucker}, {Reid}, {Jones}, {Silverman},
  {Chornock}, {Challis}, {Yuan}, {Brown}, \& {Foley}}]{Riess16}
{Riess}, A.~G., {Macri}, L.~M., {Hoffmann}, S.~L., {et~al.} 2016, \apj, 826, 56

\bibitem[{{Ripepi} {et~al.}(2019){Ripepi}, {Molinaro}, {Musella}, {Marconi},
  {Leccia}, \& {Eyer}}]{Ripepi_GDR2}
{Ripepi}, V., {Molinaro}, R., {Musella}, I., {et~al.} 2019, \aap, 625, A14

\bibitem[{{Sch{\"o}nrich} {et~al.}(2019){Sch{\"o}nrich}, {McMillan}, \&
  {Eyer}}]{Schonrich19}
{Sch{\"o}nrich}, R., {McMillan}, P., \& {Eyer}, L. 2019, \mnras, 487, 3568

\bibitem[{{Stassun} \& {Torres}(2018)}]{Stassun18}
{Stassun}, K.~G. \& {Torres}, G. 2018, \apj, 862, 61

\bibitem[{{Stassun} \& {Torres}(2021)}]{StassunTorres21}
{Stassun}, K.~G. \& {Torres}, G. 2021, \apjl, 907, L33

\bibitem[{{van Belle} {et~al.}(2020){van Belle}, {Schaefer}, {von Braun},
  {Nelan}, {Hartman}, {Boyajian}, {Lopez-Morales}, \& {Ciardi}}]{vanBelle20}
{van Belle}, G.~T., {Schaefer}, G.~H., {von Braun}, K., {et~al.} 2020, \pasp,
  132, 054201

\bibitem[{{Wilson} \& {Hilferty}(1931)}]{Wilson31}
{Wilson}, E.~B. \& {Hilferty}, M.~M. 1931, Proceedings of the National Academy
  of Science, 17, 684

\bibitem[{{Xu} {et~al.}(2019){Xu}, {Zhang}, {Reid}, {Zheng}, \& {Wang}}]{Xu19}
{Xu}, S., {Zhang}, B., {Reid}, M.~J., {Zheng}, X., \& {Wang}, G. 2019, \apj,
  875, 114

\bibitem[{{Zinn}(2021)}]{Zinn21}
{Zinn}, J.~C. 2021, \aj, 161, 214

\bibitem[{{Zinn} {et~al.}(2019){Zinn}, {Pinsonneault}, {Huber}, \&
  {Stello}}]{Zinn18}
{Zinn}, J.~C., {Pinsonneault}, M.~H., {Huber}, D., \& {Stello}, D. 2019, \apj,
  878, 136

\bibitem[{Zonca {et~al.}(2019)Zonca, Singer, Lenz, Reinecke, Rosset, Hivon, \&
  Gorski}]{HEALPy19}
Zonca, A., Singer, L., Lenz, D., {et~al.} 2019, Journal of Open Source
  Software, 4, 1298

\end{thebibliography}

\begin{appendix}

\section{Parallax difference for classical pulsators}
\label{AppCC}

\begin{figure}
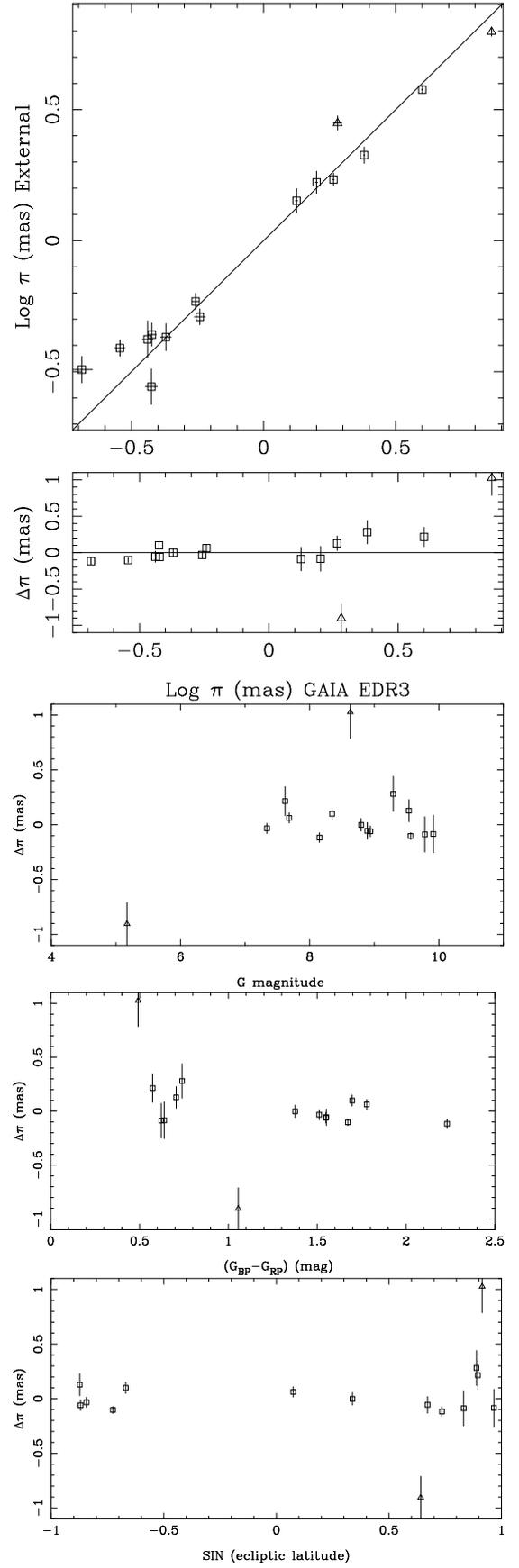

\centering

\begin{minipage}{0.39\textwidth}
\resizebox{\hsize}{!}{\includegraphics{ComparePar_15CTR.ps}}
\end{minipage}
\begin{minipage}{0.39\textwidth}
\resizebox{\hsize}{!}{\includegraphics{ParOff_Gmag_FGS_15CTR.ps}}
\end{minipage}
\begin{minipage}{0.39\textwidth}
\resizebox{\hsize}{!}{\includegraphics{ParOff_BpRp_FGS_15CTR.ps}}
\end{minipage}
\begin{minipage}{0.39\textwidth}
\resizebox{\hsize}{!}{\includegraphics{ParOff_Elat_FGS_15CTR.ps}}
\end{minipage}

\caption{Difference between the independent trigonometric parallax and the GEDR3 parallax plotted against $G$, $(G_\text{BP}-G_\text{RP})$ colour,
  and $\sin \beta$ for the 15 CCs, T2C, and RRL stars. Two outliers are plotted as open triangles.
}
\label{Fig:FGS_15CC}
\end{figure}

Figure~\ref{Fig:FGS_15CC} is as Figs.~\ref{Fig:FGS_ALL} and \ref{Fig:FGS_minus2} for the sample of 
15 CCs, T2C, and RRL stars. Two outliers are plotted as open triangles. They are FF Aql and Polaris~B.
FF Aql is very bright ($G$ = 5.1~mag) and this may be the reason for the offset.
For Polaris~B the reason is less clear. The GOF (3.55) and RUWE (1.15) easily fall within the applied selection criteria.
The difference between the FGS parallax for Polaris~B and the {\it Hipparcos} parallax for Polaris~A have been
discussed in the literature without reaching a conclusion on its implications (\citealt{Bond18,Anderson18}; appendix~B in \citealt{Gr_GDR2}).

\end{appendix}

\end{document}